\documentclass[11pt,a4paper]{article}
\usepackage{amsmath,amssymb}
\def\1#1{{\bf #1}}
\def\2#1{{\cal #1}}\def\9#1{{\sl #1}}\def\4#1{{\tt #1}}\def\5#1{{\sf #1}}
\def\6#1{{\mathfrak #1}}\def\7#1{{\mathbb #1}}\def\8#1{{\rm #1}}
\def\9#1{{\cal #1}}

\def\3{{\ss}}

\def\ol{\overline}

\def\beq{\begin{eqnarray}}
\def\eeq{\end{eqnarray}}
\def\vs{\vspace{0.2cm} \\}

\newtheorem{The}{Theorem}[section]
\newtheorem{Def}[The]{Definiton}
\newtheorem{Lem}[The]{Lemma}
\newtheorem{Pro}[The]{Proposition}
\newtheorem{Cor}[The]{Corollary}
\def\bdef{\begin{Def}\1: \em}
\def\eef{\end{Def}}

\def\blem{\begin{Lem}\1: }
\def\elem{\end{Lem}}
\def\bthe{\begin{The}\1: }
\def\ethe{\end{The}}
\def\bpro{\begin{Pro}\1: }
\def\epro{\end{Pro}}
\def\bcor{\begin{Cor}\1: }
\def\ecor{\end{Cor}}

\def\supp{{\rm supp}}

\def\id{{\rm id}}

\def\al{\alpha}
\def\be{\beta}
\def\gam{\gamma}\def\Gam{\Gamma}
\def\lam{\lambda}\def\Lam{\Lambda}
 
\def\te{\theta}

\def\sgm{\sigma}\def\Sgm{\Sigma}
\def\om{\omega}\def\Om{\Omega}

\def\bpr{\paragraph*{\it Proof.}}

\def\epr{\abs$\blacksquare$\abs}

\def\pa{\partial}
\def\<{\langle \ }
\def\>{ \ \rangle}
\def\bla{\biggl\langle \ }
\def\bra{ \ \biggr\rangle}

\def\bs{\backslash}

\def\bdes{\begin{enumerate}}
\def\edes{\end{enumerate}}
\newcommand\itno[1]{\item[{\it ({#1})}]}

\def\bmat{\left( \begin{array}{ccc} }
\def\emat{\end{array} \right)}

\def\beqa{\begin{eqnarray*}}
\def\eeqa{\end{eqnarray*}}
\def\bdia{\begin{diagram}}
\def\edia{\end{diagram}}

\def\ul{\underline}

\def\olt{\ \overline{\otimes}\ }

\def\rTo{ \ \longrightarrow \ }
\def\rMapsto{ \ \longmapsto \ }

\def\abs{\vskip 0.5cm\noindent}
\def\bcase{\left\{ \begin{array}{ccc} }
\def\ecase{\end{array} \right\}}

\title{\bf Constructive aspects of algebraic euclidean field theory}
\author{{\it Dirk Schlingemann} \\
The Erwin Schr\"odinger International Institute \\ 
for Mathematical Physics (ESI)\\
Vienna}
\begin{document}
\maketitle
\abstract{This paper is concerned with constructive 
and structural aspects  
of euclidean field theory. We present a C*-algebraic 
approach to lattice field theory. Concepts like
block spin transformations, action, effective action, and 
continuum limits are generalized and reformulated within the C*-algebraic 
setup. Our approach allows to relate to each family of lattice models  
a set of continuum limits which satisfies 
reflexion positivity and translation invariance which suggests a 
guideline for constructing euclidean field theory models.
The main purpose of the present paper is to combine the 
concepts of constructive field theory with the axiomatic 
framework of algebraic euclidean field theory in order to 
separate model independent aspects from model specific properties.}
\newpage
\tableofcontents
\newpage
\section{Introduction}
To begin with, we explain 
why euclidean field theory is of interest 
when constructive purposes are concerned. Furthermore, 
we briefly explain the basic notions which we are dealing with.   
In the second part of this section, we give an overview 
of the content of this paper by illustrating our main concepts and 
ideas.

\subsection{Motivation}
The techniques of euclidean field theory are  
powerful tools in order to construct 
quantum field theory models. 
Compared to the method of canonical quantization in Minkowski space,
which, for instance, has been used for the construction of 
$P(\phi)_2$ and Yukawa$_2$ models
\cite{GlJa1,GlJa3,GlJa4,Schra0,Schra1}, the methods 
of euclidean field theory simplify the construction of 
interactive quantum field theory models. 

The existence of the $\phi^4_3$ model as a Wightman theory 
has been established by using euclidean methods 
\cite{FeldOst,SeilSim76,MagSen76}. In the contrary  
the methods of canonical quantization are much more 
difficult to handle and lead by no means as far as euclidean techniques do. 
Only the proof of the positivity of the 
energy has been carried out within the hamiltonian framework 
\cite{GlJa1,GlJa5}. 

Motivated by the considerations above, a C*-algebraic version 
of the Osterwalder-Schrader reconstruction scheme 
has been worked out in \cite{Schl97}. 
The starting point of the analysis in \cite{Schl97} is a so called 
{\em euclidean field}. 
Within the present paper, we consider a particular class of 
euclidean fields, namely those which are statistical mechanics.
These particular euclidean fields are called 
{\em euclidean statistical mechanics}. 
We point out that within the subsequent considerations 
all physical motivations and 
interpretations are concerned with statistical mechanical systems
and not with the quantum field theory model which can be 
reconstructed from it. The axioms which we propose in \cite{Schl97}
for an euclidean field theory
are motivated by an analogous point of view as it has been 
used for the Haag-Kastler axioms \cite{HK}.  

In order to set up our language and the notions we are going to use, 
we briefly introduce and explain the mathematical formulation of 
the  concept of statistical mechanics 
from a C*-algebraic point of view. 

We apologize
for being very formal within this part of the present section, but 
one aspect of our basic philosophy is to realize the physical notions and 
concept, we are dealing with, in terms of clear mathematical objects.  

In order to describe a statistical mechanics,  
we consider a C*-algebra $\6A$ where the 
self adjoint elements describe observations related to the 
system under considerations. Each observable can be {\em localized} 
within open regions $\9U$ of a topological space $X$.
This region is related to particular {\em properties} of the 
corresponding quantity which can be measured in a certain 
experiment.  
For instance, one may think of a stochastic process, where observations
(events) can be localized within a time interval $I\subset\7R_+$, i.e.
in this case the topological space $X=\7R_+$ is simply 
the set of positive real numbers. 

As a mathematical realization of the notion 
{\em statistical mechanics}, we propose the following list of axioms:

\paragraph{\it SM1:}
Let $\9K$ be a collection  of open sets in $X$. 
The first ingredient of a statistical mechanics 
is a net of C*-subalgebras  $\ul{\6A}$
\beqa
\ul{\6A}:\9K\ni \9U\rMapsto \6A(\9U) \subset \6A
\eeqa
which is inclusion preserving, i.e.
\beqa
\9U\subset\9U_1 &\Rightarrow& {\6A}(\9U)\subset\6A(\9U_1) \ \ .
\eeqa
A region $\9U\in\9K$ can be regarded as a set of {\em properties} 
which the observables in $\6A(\9U)$ have in common. 

\paragraph{\it SM2:}
In order to describe the dynamics and symmetries of the system, we consider a 
group $G$, which acts continuously on $X$, and a
group homomorphism 
\beqa
\gam\in\8{Hom}(G,\8{Aut}\6A) 
\eeqa
from $G$ into the automorphism group of $\6A$. 
We require that $\gam$ acts partially covariantly, i.e.
\beqa
\gam_g\6A(\9U) &=&\6A(g\9U) 
\eeqa
for each $(g,\9U)\in G\times\9K$ with $g\9U\in\9K$. 

\paragraph{\it SM3:}
In addition to that, if for $\9U,\9U_1\in\9K$  
the set $\9U$ is a proper subset of $X\bs\9U_1$, then  
the algebras 
$\6A(\9U)$ and $\6A(\9U_1)$ are statistically independent 
(see \cite{Roo}). Roughly speaking, two observations which have no 
properties in common do not disturb each other.

\paragraph{\it SM4:}
Finally, we consider a state $\om$ is a state 
on the C*-algebra $\6A$ which is $G$-invariant, i.e.
$\om\circ\gam_g=\om$ for each $g\in G$.
The state $\om$ describes a basic distribution of events and 
the set of physically admissible states of the system under 
consideration is the norm closed convex hull $\9F_\om$
of the set of states
\beqa
\biggl\{\om_v:a\mapsto{\< \om, v^*av \>\over\< \om,v^*v\>}
\biggm| v\in\6A:\<\om,v^*v\>\not=0\biggr\}
\eeqa
which is called the {\em folium} generated by $\om$.
It is required that the GNS-representation of $\om$ is faithful
which is a sensible condition since, if the GNS-representation $\pi_\om$ 
is not faithful, then the ideal $\6J_\om=\pi_\om^{-1}(0)$
is irrelevant when physical aspects are concerned.
Without changing the physical content of the system under consideration 
we can replace the algebra $\6A$ by the quotient 
C*-algebra $\6A/\6J_\om$.   
\abs

The tuple $\Lam=(\ul{\6A},\gam,\om,X,G,\9K)$ which fulfills the axioms
{\it SM1} - {\it SM4} is called a {\em statistical 
mechanics}. If $\6A$ is an abelian 
C*-algebra, then we call $\Lam$ a {\em classical statistical mechanics}. 

For later purpose, it is convenient to introduce the notion of a 
{\em subsystem} of a statistical mechanics. 
A statistical mechanics  
\beqa
\Lam_1&=&(\ul{\6A}_1,\gam_1,\om_1,X_1,G_1,\9K_1)
\eeqa
is called a {\em subsystem of $\Lam$} ($\Lam_1\prec\Lam$) 
if the following conditions are fulfilled:

\paragraph{\it SU1:}
The inclusions $X_1\subset X$, $G_1\subset G$, and 
$\9K_1\subset\9K$ are valid, i.e. 
if one restricts ones considerations to a subsystem, 
then the symmetry of the underlying system can be broken.

\paragraph{\it SU2:}
The dynamics of a subsystem has to be compatible with the 
dynamics of the underlying theory. 
There exists a C*-subalgebra
$\6B\subset \6A$ and a surjective *-homomorphism 
$\rho:\6B\to\6A_1$ and  
for each $g\in G_1$ and for each $\9U\in\9K_1$ 
the following relations hold true:  
\beqa
\gam_g(\6B)&=&\6B
\vs\vs
\gam_{1,g}\circ\rho &=& \rho\circ\gam_g|_\6B 
\vs\vs 
\rho^{-1}(\6A_1(\9U)) &\subset&  \6A(\9U) \ \ .
\eeqa

\paragraph{\it SU3:}
Each state of the subsystem which is physically admissible, 
should be related to a state of the underlying theory. 
Hence one requires that 
for each state $\varphi_1\in\9F_{\om_1}$ there exists a state 
$\varphi\in\9F_\om$ such that 
\beqa
\varphi_1\circ\rho&=&\varphi|_{\6B} \ \ .
\eeqa
\abs
Two statistical mechanics 
$\Lam,\Lam_1$ are {\em equivalent} if $\Lam_1$ is a subsystem of 
$\Lam$ and vice versa. 

In general, the *-homomorphism $\rho$  
is not faithful, which can be interpreted in physical terms: 
Relations between observables within the subsystem 
are tested by states in $\9F_{\om_1}$. Within the 
underlying theory a larger set of states $\9F_\om$ can be prepared
and therefore relations between observables, which hold for the subsystem, 
can be violated within the underlying one. 

It is clear that to each localizing region $\9U\in\9K$ we can
assign a subsystem in a natural manner, namely
\beqa
\Lam_{\9U}
&:=&(\ul{\6A}_{\9K(\9U)},\gam|_{G(\9U)},\om|_{\6A(\9U)},\9U,G(\9U),\9K(\9U))
\ \prec \ \Lam
\eeqa
where $G(\9U)\subset G$ is the stabilizer subgroup of $\9U$ and 
$\9K(\9U)$ contains all sets $\9U_1\in\9K$ with $\9U_1\subset\9U$.

We are now prepared to introduce the notion of euclidean statistical mechanics.
Let $\9K^d$ be the set of open bounded convex subsets of $\7R^d$.
A {\em euclidean statistical 
mechanics} is a statistical mechanics
\beqa
(\ul{\6A},\al,\om,\7R^d,\8E(d),\9K^d)
\eeqa
where the state $\om$ fulfills the axioms:

\paragraph{\it E1:}
The state $\om$ is euclidean invariant, i.e. $\om\circ\al=\om$.

\paragraph{\it E2:}
The state $\om$ is reflexion positive: 
Let $e\in S^{d-1}$ be an euclidean time-direction and let 
$\Sgm_e$ be the hyper-plane which is orthogonal to $e$. 
The euclidean time reflexion $\te_e:\7R^d\to\7R^d$ is the reflexion
\beqa
x\rMapsto \te_e(x)&=& x-2(e\cdot x) \ e
\eeqa
where $y\cdot x$ is the canonical scalar product in $\7R^d$. 
We consider the anti-linear involution 
\beqa
j_e&:=&\al_{\te_e}\circ ^*\in\8{Aut}\6A 
\eeqa
and we require that 
\beqa
\<\om ,j_e(a)a\> &\geq&0
\eeqa
for each $a\in \6A(\7R_+e+\Sgm_e)$.

\paragraph{\it E3:}
The state $\om$ fulfills a regularity condition, namely for each 
$a,b,c\in\6A$ the map
\beqa
g\rMapsto \<\om,a\al_g(b)c \>
\eeqa
is continuous. 
\abs

By considering euclidean statistical mechanics,
the property {\it SM2} is then called {\em euclidean covariance} 
and the statistical independence in {\it SM3} is called 
{\em locality} \cite{Schl97}.

The problem of constructing non-trivial examples which fulfill 
the axioms {\it E1}-{\it E3} is rather difficult to handle. 
Up to now, the known examples for euclidean field theory models 
which are {\em not} related to free field theory models 
are examples in $d<4$ space-time dimensions.
The question whether there are interesting models in $d\geq 4$ 
dimensions is still open. 

One possible procedure, which is 
often used within the framework of constructive 
field theory, is to start from a family of lattice field theory models
which can be regarded as statistical mechanics in our sense
(see \cite{FernFrohSok92} and references given there). 
As a tool to control the 
continuum limit,
block spin transformations 
are used to relate models, which belong to a given lattice, with 
models on a finer lattices. 
This method has been applied to
scalar field theories \cite{GawKup} as well as to the 
treatment of gauge theories \cite{Bal84a}, for instance. 
But even if a suitable continuum limit exists in the sense of 
\cite{FernFrohSok92,GawKup,Bal84a}, then this    
does not imply that the axioms {\it E1}-{\it E3} are fulfilled. 
Since one works here with cubic lattices, it is extremely difficult 
to prove the rotation invariance of the model which is indeed a 
crucial property for passing from a euclidean field theory to 
a quantum field theory in Minkowski space. One nice idea, which 
works at least in $d=2$ dimensions and which makes use of the 
facts developed in \cite{Bal84a}, is presented in \cite{King86}.
We also refer the reader to \cite{FernFrohSok92} where this problem is also 
mentioned. 

Within this paper we also work with cubic lattices and 
the problem of rotation invariance is discussed 
within a forthcoming paper. Concerned with this simplification,  
we study statistical mechanics 
\beqa
\Lam&=&(\ul{\6A},\al,\om,\7R^d,\7Q_b^d,\9K^d) \ \ ,
\eeqa
with $\7Q_b^d:=\cup_n b^{-n}\7Z^d$, $b\in\7N$,  
i.e. 
the net $\ul{\6A}$ 
is translationally covariant with respect to a dense subgroup 
$\7Q_b^d\subset\7R^d$ of rational translations.  
The axioms {\it E1}-{\it E3} for the state $\om$ are also substituted by 
weaker properties {\it WE1}-{\it WE3}:

\paragraph{\it WE1:}
The state $\om$ is translationally invariant, i.e. $\om\circ\al=\om$.

\paragraph{\it WE2:}
The state $\om$ is reflexion positive with respect to the 
directions $e_k$, where $e_k$ is the unit vector with 
components $(e_k)_l=\delta_{kl}$.

\paragraph{\it WE3:}
The state $\om$ fulfills a regularity condition, namely for each 
$a,b,c\in\6A$ the map
\beqa
\7Q_b^d\ni g\rMapsto \<\om,a\al_g(b)c \>
\eeqa
is continuous. 
\abs

We call the tuple $\Lam$ a  
{\em weak euclidean statistical mechanics} if 
it satisfies the axioms {\it WE1}-{\it WE3} and the pair 
$(\ul{\6A},\al)$ is called a {\em weak euclidean net of C*-algebras}. 

We expect that the axioms for a weak euclidean statistical mechanics 
are not sufficient to 
construct a Haag-Kastler net within a vacuum representation 
from these data. Nevertheless, a weak euclidean statistical mechanics can be 
treated as a physical system by its 
own right.
\subsection{Overview}
After we have introduced the general concepts and notations in the 
previous section, we outline here the basic 
ideas and concepts which are 
developed within this paper in a concrete manner. 

We consider the lattice of the discretized torus 
$\Sgm_0(n)=b^{-n^0}\7Z^d/b^{n^1}\7Z^d$ where 
$b\in\7N$ is odd and $n=(n^0,n^1)\in\7Z^2$ is a pair of integer numbers. 
The corresponding sets of $q$-cubes are denoted by 
$\Sgm_q(n)$, $q\leq d$. The set of $q$-cubes of the dual lattice 
is denoted by $\Sgm^*_q(n)$ and we use the symbol $*$ for 
the isomorphism which maps $\Sgm^*_{d-q}(n)$ onto $\Sgm_q(n)$ and 
vice versa. We introduce a partial ordering on 
$\7Z^2$: We write $n\prec n_1$ for $n^j\leq n_1^j$, $j=0,1$.

For a given lattice, we 
build the C*-algebra of bounded continuous functions 
\footnote{For a C*-algebra
$A$ we write 
$\6A_{(n,X[A])}=\6A_{(n,A)}:=
\otimes_{\Delta\in\Sgm_d(n)}\{\Delta\}\times A$ 
where $X[A]$ denotes the spectrum of $A$.}
\beqa
\6A_{(n,\7R)}&:=&\9C_\8b(\7R^{\Sgm_d(n)}) \ \ .
\eeqa
The algebra $\6A_{(n,\7R)}$ contains subalgebras $\6A_{(n,\7R)}(\9U)$
which are related to an open convex sets 
$\9U\subset [-b^{n^1},b^{n^1}]^d$ in euclidean space, namely 
a function $a\in\6A_{(n,\7R)}$ is localized in $\9U$ if it only depends upon 
the variables $u(\Delta),\phi(\Delta)\subset \9U$, where 
$\phi$ is an appropriate chart from the torus 
$\7R^d/b^{n^1}\7Z^d$ into $\7R^d$.

As an example for a lattice field theory model 
we consider a lattice  action functional of the form 
\beq
\1s_{(\lam,n)}(u)
&=&\lam_0(n)\sum_{\Gam\in\Sgm_{d-1}(n)} *\8d * u (\Gam) 
\nonumber\vs\nonumber\vs
&+&
\sum_{\Delta\in\Sgm_d(n)} \sum_{l=1}^L \lam_l(n)u(\Delta)^{2l}  
\label{scmo}
\eeq
which induces a state $\eta_{(\lam,n)}$ on $\6A_{(n,\7R)}$ by defining
\beqa
\< \eta_{(\lam,n)},a \>&=&
\1z_{(\lam,n)}^{-1}\int \8du \ \exp(-\1s_{(\lam,n)}(u)) \ a(u)
\eeqa
where the partition function $\1z_{(\lam,n)}$ is for normalization.
In order to formulate the important properties of the states $\eta_{(\lam,n)}$,
we look at particular automorphisms on $\6A_{(n,\7R)}$. 
The group $b^{-n^0}\7Z^d$ acts on the set of cubes 
$\Sgm_d(n)$ in a natural manner and for each $g\in b^{-n^0}\7Z^d$ 
we introduce an automorphism $\be_{(n,g)}$ on $\6A_{(n,\7R)}$ 
by the prescription
\beqa
\be_{(n,g)}a(u)&:=& a(u\circ g) \ \ .
\eeqa
Moreover, 
the euclidean time reflexions $\te_\mu=\te_{e_\mu}$, $\mu=1\cdots d$, 
also act on  $\Sgm_d(n)$ and we define anti-automorphisms $j_\mu$ 
\beqa
j_{(n,\mu)} a(u)&:=&\bar a(u\circ \te_\mu) \  \ .
\eeqa
It can be proven that the states $\eta_{(\lam,n)}$ are invariant 
under the automorphisms $\be_{(n,g)}$ and that they are 
reflexion positive, i.e.
\beqa
\<\eta_{(\lam,n)},j_{(n,\mu)}(a)a\> &\geq& 0
\eeqa 
for each operator $a$ which is localized in $\7R_+e_\mu+\Sgm_{e_\mu}$.

Let $\9K^d_n$ be the collection of all open convex sets in
$[-b^{n^1},b^{n^1}]^d$ and let $\6J_{(n,\eta)}$ be the 
kernel of the GNS-representation of $\eta_n$. The prescription 
\beqa
\ul{\6A}_{(n,\7R|\eta)}:\9K^d_n\ni \9U\rMapsto
\6A_{(n,\7R|\eta)}(\9U) \ := \ \6A_{(n,\7R)}(\9U)/\6J_{(n,\eta)}
\eeqa
yields a concrete example for a classical statistical mechanics,
namely the tuple 
\beqa
\Lam_n&=&(\ul{\6A}_{(n,\7R|\eta)},\be_n,\eta_n,
\7R^d,b^{-n^0}\7Z^d,\9K^d_n)  \ \ .
\eeqa
In the subsequent, we call $\Lam_n$ 
a {\em lattice field theory} if the state $\eta_n$ is 
$b^{-n^0}\7Z^d$ invariant and reflexion positive.

\paragraph{\it Continuum limits for lattice field theories:}
As already mentioned,  
in order to control the continuum limit of lattice field theory models 
the concept of block spin
transformations turned out to be a useful tool. For a review of the 
basic ideas, we refer the reader to \cite{FernFrohSok92} and references
given there.
We reformulate the basic concepts of block spin transformations 
from a C*-algebraic point of view. Each  
configuration $u\in\7R^{\Sgm_d(n+k)}$, $k\in\7N^2$, can be
identified with a configuration 
$p_{(n,n+k)}u\in\7R^{\Sgm_d(n)}$ by an averaging procedure. 
Usually, the averaging map $p_{(n,n+k)}$ is defined by 
the block average
\beq\label{bs1}
(p_{(n,n+k)}u)(\Delta_0)&:=&b^{-dk^0}\sum_{\Delta\subset\Delta_0} u(\Delta)
\ \ .
\eeq
A simplified version of a block spin transformation can by 
obtained by setting 
\beq\label{bs2}
(p_{(n,n+k)}u)(\Delta_0)&:=&u(\Delta_{(n,n+k|\Delta_0)})
\eeq
where $\Delta_{(n,n+k|\Delta_0)}$ is the unique cube contained in 
$\Delta_0$ which contains the point $*\Delta_0$ in the 
dual lattice.
 
The block spin transformations can be used to identify 
operators in $\6A_{(n,\7R)}$ with operators in $\6A_{(n+k,\7R)}$, namely 
\beqa
\iota_{(n+k,n)}a&:=&a\circ p_{(n,n+k)}
\eeqa
defines a faithful *-homomorphism form $\6A_{(n,\7R)}$ 
into $\6A_{(n+k,\7R)}$. 
In contrary to the common literature, we 
distinguish here between {\em block spin} transformations
and {\em renormalization group} transformations. One important feature 
of block spin transformations is that localizing regions are preserved,
i.e. $\iota_{(n+k,n)}\6A_{(n,\7R)}(\9U)\subset\6A_{(n+k,\7R)}(\9U)$.
Hence there is no scaling involved as block spin transformations
are concerned. On the other hand, renormalization group 
transformations identify operators which are localized in $\9U$  
with operators, localized in a scaled region $\lam\9U$.
An overview of the basic ideas of renormalization group transformations
applied to constructive field theory can be found in   
\cite{GawKup,BrDiHu97} and references given there. 
The general concept of 
renormalization group transformations from an axiomatic point of view 
is presented in \cite{BuVer95,BuVer97} and related work. 

By looking at algebraic properties, in Section \ref{su2} the 
general concept of block spin transformation is introduced 
within the C*-algebraic setting. 
As we shall describe in Section \ref{su3}, by performing the 
C*-inductive limit, one constructs from a given 
family of block spin transformations 
$\iota=(\iota_{(n,n_0)})_{n_0\prec n}$ and from 
the lattice algebras $\6A_{(n,\7R)}$ a C*-algebra $\6A_{(\iota,\7R)}$ which 
can be regarded as the C*-algebra for the continuum model.  
One obtains a net of C*-algebras
\beqa
\ul{\6A}_{(\iota,\7R)}:\9U\rTo\6A_{(\iota,\7R)}(\9U)
\eeqa
on which the dense subgroup $\7Q^d_b=\cup_{n\in\7N}b^{-n}\7Z^d\subset\7R^d$
acts covariantly by automorphisms $\be_{(\iota,g)}$ and thus this yields 
a weak euclidean net of C*-algebras $(\ul{\6A}_{(\iota,\7R)},\be_\iota)$.

One aim of this paper is to analyze the space of $\7Q^d_b$-invariant and 
reflexion positive states $\6S_{(\iota,\7R)}$ 
on $\6A_{(\iota,\7R)}$. 
The application of block spin transformations to states
leads to a net of invariant reflexion positive states 
\beqa
(\eta_{n+k}\circ\iota_{(n+k,n)})_{k\in\7N^2}
\eeqa
on $\6A_{(n,\7R)}$ which has, according to compactness arguments, 
weak limit points. We denote this 
weak limit points by $\varphi_n:=\1E[\xi\otimes\eta]_n$, where 
$\xi$ labels a limit point, more precisely, 
$\xi$ is a measure on the space ${\bar\7Z}^2\bs\7Z^2$, where 
${\bar\7Z}^2$ is the spectrum of the C*-algebra of bounded 
functions on $\7Z^2$. The consistency condition 
\beqa
\varphi_{n+k}\circ\iota_{(n+k,n)}&=&\varphi_n
\eeqa
is fulfilled and hence there is a unique state $\varphi\in\6S_{(\iota,\7R)}$ 
on the C*-inductive limit $\6A_{(\iota,\7R)}$ such that
\beqa
\varphi\circ\iota_n&=&\varphi_n
\eeqa
where $\iota_n$ is the embedding of $\6A_{(n,\7R)}$ into
$\6A_{(\iota,\7R)}$. For a given family of lattice field theory models
\beqa
(\ul{\6A}_{(n,\7R|\eta)},\be_n,\eta_n,\7R^d,b^{-n^0}\7Z^d,\9K_n^d)_{n\in\7Z^2} 
\eeqa
we symbolize the corresponding set of 
{\em continuum limits} by $\6S_{(\iota,\7R)}[\eta]\subset\6S_{(\iota,\7R)}$.
Each continuum limit $\varphi\in\6S_{(\iota,\7R)}[\eta]$ gives 
rise to a classical statistical mechanics   
\beqa
\Lam&=&(\ul{\6A}_{(\iota,\7R|\varphi)},\be_\iota,\varphi,\7R^d,\7Q_b^d,\9K^d)
\eeqa
where the net $\ul{\6A}_{(\iota,\7R|\varphi)}$ is given by 
\beqa
\6A_{(\iota,\7R|\varphi)}:\9U\rMapsto\6A_{(\iota,\7R)}(\9U)/
\6J_{(\iota,\varphi)}
\eeqa
and $\6J_{(\iota,\varphi)}$ is the kernel of the GNS-representation of 
$\varphi$.

The self adjoint operators in 
$\6A_{(\iota,\7R|\varphi)}$ correspond to observations with respect to the 
full energy momentum range. 
By setting $\varphi_n:=\varphi\circ\iota_n$, each lattice field theory 
\beqa 
\Lam_n&=&(\ul{\6A}_{(n,\7R|\varphi)},\be_n,\varphi_n,\7R^d,
b^{-n^0}\7Z^d,\9K_n^d)
\eeqa
is a proper subsystem of $\Lam$
which corresponds to observations within the energy momentum 
range $[b^{-n^1},b^{n^0}]$ and we regard $\Lam_n$ as 
an {\em effective theory} in which observations within the 
energy momentum range $[0,b^{-n^1}]\cup [b^{n^0},\infty)$ are 
not admissible. The $\7Q_b^d$ covariance of the 
effective theory is broken and only the $b^{-n^0}\7Z^d$
covariance remains. 

At this point, 
we have to emphasize that our considerations essentially rely on the 
C*-algebraic point of view. 
The advantage in comparison to 
non-C*-based approaches (see for example \cite{FernFrohSok92}) 
is that we always get continuum limits
no matter how our input data 
$\eta=(\eta_n)_{n\in\7Z^2}$ are chosen.  In particular, 
by looking at the family $\eta_\lam=(\eta_{(\lam,n)})_{n\in\7Z^2}$
of scalar field theory models, given by Equation (\ref{scmo}), 
we get continuum limits for arbitrary couplings $(\lam_l(n))_{n\in\7Z^2}$,
$l=0\cdots L$. Even in case of a perturbatively non-renormalizable 
model, it makes sense to study the set of continuum limits.  

On the other hand, the fact that there are weak limit points is not 
sufficient for concluding the existence of interesting models. 
Therefore, the problem which occur here 
is to get detailed information about the states in
$\6S_{(\iota,\7R)}[\eta]$. At this point, we introduce 
a rough classification of families of states 
by considering the possible 
limit points of a given family $\eta$. 
\bdes

\itno 1
For a given family $\eta$ 
every limit point in $\6S_{(\iota,\7R)}[\eta]$ is a character which is 
the most trivial case.  

\itno 2
There is another uninteresting case, namely
each state in $\6S_{(\iota,\7R)}[\eta]$ is ultra local, i.e.
each state $\varphi\in\6S_{(\iota,\7R)}[\eta]$ has 
no correlation for two operators $a_j\in\6A_{(\iota,\7R)}(\9U_j)$, 
$j=1,2$, which are 
localized in disjoint regions $\9U_1\cap\9U_2=\emptyset$:
\beqa
\<\varphi,a_1a_2\> &=& \<\varphi_1,a_1\>\<\varphi_2,a_2\>
\eeqa
for suitable states $\varphi_j$ on $\6A_{(\iota,\7R)}(\9U_j)$, $j=1,2$.
This implies that, if the corresponding theory in 
Minkowski space exists, then it is the constant field.
The notion of ultra local (scalar) fields is explained  
in \cite{Klau70}. In particular an application of the 
measures, constructed in \cite{AshLew95}, to euclidean field theory leads to 
ultra local models.

\itno 3 
There exists a limit point $\varphi\in \6S_{(\iota,\7R)}[\eta]$ which is not 
ultra local. 
\edes

By looking at our example of scalar fields, the case 
{\it (3)} can be subdivided into two further cases:
\bdes
\itno {3.1}
Let $\varphi\in \6S_{(\iota,\7R)}[\eta]$ be a non-ultra local state, then
it is equivalent to a gaussian state.

\itno {3.2}
There exists a limit point $\varphi\in \6S_{(\iota,\7R)}[\eta]$ which is not
ultra local and which is not equivalent to a gaussian state.
\edes

We have to mention at this point that for many examples 
case {\it (1)} can be excluded. One now asks the following question:

\paragraph{\it Question:}
Can we decide, by studying the family of states $\eta$, whether the 
case {\it (3)} is valid or not? 

\abs

In order to show the existence of states in $\6S_{(\iota,\7R)}$,
which are not ultra local, we propose the 
following strategy: For a cube $\Delta\in\Sgm_d(n)$ and for an
operator $a\in \9C_\8b(\7R)$ we define the function 
$\Phi_n(\Delta,a)$ by 
\beqa
\Phi_n(\Delta,a)(u)&:=&a(u(\Delta)) \ \ .
\eeqa
Let $\Delta_1,\Delta_2\in\Sgm_d(n)$ be two disjoint 
cubes $\Delta_1\cap\Delta_2=\emptyset$. 
Find a continuous bounded positive function $h\in\9C_\8b(\7R)$ and 
a family of states $\eta$ such that there exists a constants 
$c^\pm_{(n,h,\Delta_1,\Delta_2)}>0$ 
with 
\beqa
c^+_{(n,h,\Delta_1,\Delta_2)}&\geq&
|\<\1c_{[\eta_{n+k}\circ\iota_{(n+k,n)}]},\Phi_n(\Delta_1,h)\otimes
\Phi_n(\Delta_2,h)\>|
\vs\vs
&>&c^-_{(n,h,\Delta_1,\Delta_2)}
\eeqa
for large $k$. Here we define for any state $\om$ its correlation
by
\beqa
\<\1c_{[\om]},a\otimes b\>&:=&\<\om,ab\> \ - \ \<\om,a\> \<\om,b\> \ \ .
\eeqa
Since the bound is uniform in $k$, 
there exists a state $\varphi\in\6S_{(\iota,\7R)}[\eta]$ such that the 
correlation of $\varphi_n=\varphi\circ\iota_n$ fulfills  
the bounds 
\beqa
c^+_{(n,h,\Delta_1,\Delta_2)}&\geq&
|\<\1c_{[\varphi_n]},\Phi_n(\Delta_1,h)\otimes\Phi_n(\Delta_2,h)\>|
\vs\vs
&>&c^-_{(n,h,\Delta_1,\Delta_2)}
\eeqa
which implies that $\varphi$ is not ultra local. 
From the invariance properties of $\varphi_n$ we conclude that 
this bound holds for each pair of cubes which can be obtained by 
applying a transformation $g\in b^{-n^0}\7Z^d$ to $(\Delta_1,\Delta_2)$.
Hence the constant $c_{(n,\Delta_1,\Delta_2)}$ 
only depends on the orbit of $(\Delta_1,\Delta_2)$
under the action of $b^{-n^0}\7Z^d$. Let $d(\Delta_1,\Delta_2)$ 
be the distance of the cubes $(\Delta_1,\Delta_2)$ and let us assume 
that the upper bound $c^+_{(n,h,\Delta_1,\Delta_2)}$ has the form
\beqa
c^+_{(n,h,\Delta_1,\Delta_2)}&=&K_{(n,h)}
\exp\biggl(- {d(\Delta_1,\Delta_2)\over\ell(n,h)}\biggr)
\eeqa
with two constants $K_{(n,h)},\ell(n,h)$, then the constant 
$\ell(n,h)$ plays the role of the correlation length.
A proposal how to tackle the problem of estimating 
correlations is given in Appendix \ref{app1}.

\paragraph{\it Action, effective action, and continuum limits:}
We assume now that somebody has already
constructed a weak euclidean statistical mechanics 
\beqa
\Lam&=&(\ul{\6A}_{(\iota,\7R|\om)},\be_\iota,\om,\7R^d,\7Q^d_b,\9K^d) \ \ .
\eeqa
Then it is natural to ask whether one can construct 
new theories out of $\Lam$ by a suitable deformation procedure.
Remember that $\6A_{(\iota,\7R|\om)}$ denotes the C*-algebra
\beqa
\6A_{(\iota,\7R|\om)}&:=&\6A_{(\iota,\7R)}/\6J_{(\iota,\om)}
\eeqa
where $\6J_{(\iota,\om)}$ is the kernel of the GNS-representation 
of $\om\in\6S_{(\iota,\7R)}$. The basic idea is to perturb 
each of the subsystems 
\beqa
\Lam_n&=&(\ul{\6A}_{(n,\7R|\om)},\be_n,\om_n,\7R^d,b^{-n^0}\7Z^d,\9K_n^d)
\eeqa
separately, by replacing each of the states $\om_n$ by 
appropriate states $\eta_n\in\9F_{\om_n}$. If we assume that 
$\eta^{(k)}_n:=\eta_{n+k}\circ\iota_{(n+k,n)}$ is contained in $\9F_{\om_n}$
for each $k\in\7N^2$, then we obtain for each 
$n\in\7Z^2$ and for each $k\in\7N^2$ a subsystem 
\beqa
\Lam_n^{(k)}
&:=&(\ul{\6A}_{(n,\7R|\eta^{(k)}_n)},\be_n,\eta^{(k)}_n,\7R^d,
b^{-n^0}\7Z^d,\9K_n^d)\ \ \prec \ \ \Lam_n \ \ .
\eeqa   
which is, in particular, a subsystem of $\Lam$. There are also 
examples for which the theories $\Lam_n^{(k)}\cong\Lam_n$ are equivalent
for each $k$. 
Formally, the relation $\Lam_n^{(k)}\cong\Lam$ may be no  
longer valid in the continuum limit $k,n\to\infty$. More precisely,
for a continuum limit $\varphi\in\6S_{(\iota,\7R)}[\eta]$ 
the corresponding theory
\beqa
\Lam^{(\varphi)}
&=&(\ul{\6A}_{(n,\7R|\varphi)},\be_\iota,\varphi,\7R^d,\7Q^d_b,\9K^d)
\eeqa
is, however, {\em not equivalent} to the theory where we have 
started from.

But one may ask whether the subsystem 
$\Lam^{(\varphi)}_n$, which corresponds to the energy momentum range 
$[b^{-n^1},b^{n^0}]$, is a subsystem of $\Lam_n$.
This question is related to the existence of an effective action \cite{GawKup}.
The states $\eta_n$ under consideration are of the form
\beqa
\<\eta_n,a\>&=&\int_{\7R^{\Sgm_d(n)}}\8d\om_n(u) \ \1v_n(u) \ a(u)
\eeqa
and we call 
the family of functions $\1v=(\1v_n)_{n\in\7Z^2}$ an {\em action}.

Our notion of action is slightly different to the 
one which one usually finds in the literature where 
in comparison the negative logarithm 
$-\ln\1v_n$ is usually called the action.
In order to distinguish these notions we call 
$-\ln\1v_n$ the {\em action functional} with respect to $n$.
For example, choose $\om_n$ to be the gaussian part and 
$\1v_n$ to be the interaction part (see \cite{BrDiHu97}).
Within our analysis, we also consider examples where 
$\om_n$ is an ultra local state and 
$\1v_n$ contains the next neighbor coupling.

From a given action $\1v$, we obtain a new family of functions by 
\beqa
\1e_{(\om)}^{(k)}(\1v)_n(u)&:=&
\int \8d\om_{(n+k)}(u') \ \1k_{(\om,n,n+k)}(u,u') \ \1v_{n+k}(u')
\eeqa
where the kernel $\1k_{(\om,n,n+k)}(u,u')$ is determined by the 
condition
\beqa
\int \8d\om_n(u) \ \8d\om_{(n+k)}(u') \ \1k_{(\om,n,n+k)}(u,u') \ a(u')
\vs\vs
= \ \ 
\int \8d\om_{(n+k)}(u') \ a(u') \ \ .
\eeqa
We call $\1e^{(k)}_{(\om)}(\1v)$ the {\em effective action}
with respect to the action $\1v$. 
For a fixed cut-off $n\in\7Z^2$ the operation of 
$\1e_{(\om)}^{(k)}$ corresponds 
to a substitution by the underlying lattice theory on 
$\Sgm_d(n)$ by a lattice theory, also defined on 
$\Sgm_d(n)$, arising from a lattice theory 
on $\Sgm_d(n+k)$ by integrating out the corresponding high energy 
degrees of freedom (See \cite{GawKup}). 

In Section \ref{s2}, we discuss in a more general context 
conditions for $\1v$ under which there exists a family of 
measurable functions
$\1v'=(\1v'_n)_{n\in\7Z^2}$ ($\1v'_n$ is $\om_n$-measurable) such that 
\beq\label{eq1}
\< \varphi,\iota_n(a)\>&=&
\1z_{(\om,\1v',n)}^{-1}\int \8d\om_n(u)   \ \1v'_n(u) \ a(u) 
\eeq
holds for a continuum limit $\varphi\in\6S_{(\iota,\7R)}[\eta]$.
In this case, $\Lam^{(\varphi)}_n$ is a subsystem of $\Lam_n$ since 
the folium $\9F_{\varphi_n}$ is contained in 
$\9F_{\om_n}$.

In Section \ref{su8} we formulate a sufficient condition 
({\em multiplicative renormalizability})
for an action $\1v$ which allows to  construct a new action $\1v'$ 
from $\1v$ such that $\1v'$ satisfies the fix point equation 
$\1e^{(k)}_{(\om)}(\1v')=\1v'$ and therefore Equation (\ref{eq1})
(Proposition \ref{themultren}).
We have to emphasize here that the existence of $\1v'$ 
does not exclude the case {\it (2)} of ultra locality.
In order to conclude that one deals with an interesting model
one has to study $\1v'$ in more detail.  

To illustrate the notion {\em multiplicative renormalizability}, 
an ultra local example for a 
multiplicatively renormalizable action is also presented in 
Section \ref{su8} and Appendix \ref{app2} deals with a larger class of 
examples.

\paragraph{\it A large variety of lattice models:}
The C*-algebraic point of view suggests to study a large class
of lattice field theories among which there are examples which are  
rather different from 
the usual lattice field theory models, like $P(\phi)_d$ for instance. 
To some extend they can be regarded as generalized spin models. 

The abelian C*-algebra $\9C_\8b(\7R)$, which we have used 
for illustration in the previous paragraphs, can easily be replaced by  
any C*-algebra $A$ in particular by a $\sgm$-finite von Neumann algebra $M$
acting on a Hilbert space $K$. As usual, we denote by $M'$ the 
commutant of $M$, i.e. the set of bounded operators on $K$ which 
commute with all operators in $M$. 

As input data for the construction of lattice models we 
choose 
\bdes
\itno 1
a von Neumann algebra $M$, acting on a Hilbert space $K$, and 
a vector $\Om$, which is cyclic and separating for $M$, 
\itno 2
a family of positive operators $\1w\in (M'\olt M')^{\7Z^2}$
such that 
\beqa
[\1w_n\otimes\11,\11\otimes\1w_n]=0
\eeqa 
for each $n\in\7Z^2$.
\edes 

The algebra $\6A_{(n,M)}=\olt_{\Delta\in\Sgm_d(n)}\{\Delta\}\times M$
is simply the von Neumann tensor product of $M$ over $\Sgm_d(n)$ and 
the vector 
$\Om_n:=\otimes_{\Delta\in\Sgm_d(n)}\{\Delta,\Om\}$ is cyclic
and separating for $\6A_{(n,M)}$. For a cube $\Delta\in\Sgm_d(n)$ and 
an operator $a\in M$ we denote by $\Phi_n(\Delta,a)$  
operator in $\6A_{(n,M)}$  which is a tensor product of 
operators in $M$ where at $\Delta$ the factor $a$ appears and 
the unit $\11$ else. 
For a hypercube  $\Gam\in\Sgm_{d-1}(n)$
there are two unique cubes $\Delta_0,\Delta_1$ such that 
$\Delta_0\cap\Delta_1=\Gam$ and we put
$\Phi_n(\Gam,a\otimes b):=\Phi_n(\Delta_0,a)\Phi_n(\Delta_1,b)$.
We introduce a state $\eta_n$ on $\6A_{(n,M)}$ by
\beqa
\<\eta_n,a\>&:=&\1z_{(\Om,n,\1w)}^{-1}
\bla\Om_n,\prod_{\Gam\in\Sgm_{d-1}(n)}\Phi_n(\Gam,\1w_n) \ a \ \Om_n\bra
\eeqa
where the partition function $\1z_{(\Om,n,\1w)}$ is for normalization.
If the vector 
\beqa
\Psi_{(\Om,n,\1w)}&:=&\prod_{\Gam\in\Sgm_{d-1}(n)}\Phi_n(\Gam,\1w_n)^{1/2}\Om_n
\eeqa
is cyclic and separating for $\6A_{(n,M)}$ for each $n\in\7Z^2$, then 
$\eta_n$ is faithful and we obtain for each $n\in\7Z^2$ a 
lattice field theory model
\beqa
\Lam_n&:=&(\ul{\6A}_{(n,M)},\be_\iota,\eta_n,\7R^d,b^{-n^0}\7Z^d,\9K^d_n) 
\ \ .
\eeqa

For each $n\in\7N$, we consider the state  $\om_n=\<\Om_n,(\cdot)\Om_n\>$ and 
by an appropriate choice of block spin transformations $\iota$
the consistency condition
$\om_{n+k}\circ\iota_{(n+k,n)}=\om_n$ is fulfilled and we obtain 
the corresponding continuum model
\beqa
(\ul{\6A}_{(\iota,M)},\be_\iota,\om,\7R^d,\7Q_b^d,\9K^d) \ \ .
\eeqa
The state $\eta_n$ is a perturbation of $\om_n$ where the 
action is given by
\beqa
\1v:n\rMapsto  \prod_{\Gam\in\Sgm_{d-1}(n)}\Phi_n(\Gam,\1w_n) \ \ .
\eeqa
Each of the operators $\Phi_n(\Gam,\1w_n)$ induces a coupling of the 
two next neighbor cubes which have the face $\Gam$ in common. 
If $\1w_n$ is of the form $\1w_n=\1h_n\otimes\1h_n$ 
the cubes are decoupled and the resulting theory 
is ultra local. The simplest non-trivial choice for $\1w_n$ is 
$\11+\1h_n\otimes\1h_n$ for instance.  
More general, one can choose $\1w$ in the following manner:
Put  
\beqa
\1w_n&:=& \int_0^1 \1h_n(s)\otimes\1h_n(s)
\eeqa
where $\1h_n\in\9C^\infty([0,1],M')$ is a smooth function with $\1h_n(s)>0$
and \newline $[\1h(s_1),\1h(s_2)]=0$. If $\1h_n(s)=\1h_n$ is constant, then we 
would end up with an ultra local theory. Therefore we have to require 
that the derivative of $\1h_n$ does not vanish. 
For this kind of examples, the effective 
action $\1e^{(k)}_{(\om)}(\1v)$ can be computed quite explicitly
and our hope is that the corresponding continuum limits 
could be easier controlled than the continuum limits 
for $P(\phi)_d$-like models, for instance. 

There is a further nice feature of models which correspond to such actions 
like $\1v$. Particular correlation functions can be interpreted in terms of 
correlation functions of a 
different, some kind of dual, lattice field theory.
In order to explain this, we introduce for a cube $\Delta\in\Sgm_d(n)$ 
and for each $s\in [0,1]^{\Sgm_{d-1}(n)}$ a normal state on $M$:
\beqa
\<\1E^{(s)}_{(\1h,n|\Delta)}, a \>&:=&[\1z^{(s)}_{(\1h,n|\Delta)}]^{-1}
\bla\Om, \ \prod_{\Gam\in\pa\Delta} \1h_n(s(\Gam)) \ a \ \Om\bra 
\eeqa
where $\1z_{(h,n|\Delta)}(s)$ is for normalization.
This yields a state $\hat\eta_{(n,h)}$ on the algebra
$\hat\6A_{(n,[0,1])}:=\9C([0,1]^{\Sgm_{d-1}(n)})$ such
that for operators 
$(a_j)_{j=1\cdots k}$, $a_j\in M$,  
the correlation functions fulfill the relation
\beqa
\bla\eta_n,\prod_{j=1}^k \Phi_n(\Delta_j,a_j) \bra
&=&
\int\8d\hat\eta_n(s) \prod_{j=1}^k\<\1E^{(s)}_{(n,\1h|\Delta_j)},a_j\>  \ \ .
\eeqa
The state $\hat\eta_{(h,n)}$ is given by 
\beqa
\<\hat\eta_{(h,n)},\hat a\>
&=&
\1z_{(h,n)}^{-1}
\int \prod_{\Gam\in\Sgm_{d-1}(n)} \8ds(\Gam) \ \hat\1v_n(s) \ \hat a(s)
\eeqa
and $\hat\1v_n$ is given by 
\beqa
\hat\1v_n(s)&=&\prod_{\Delta\in\Sgm_d(n)} \1z^{(s)}_{(h,n|\Delta)} \ \ .
\eeqa
Indeed, particular correlation functions of the model, 
which is given by the action $\1v$, can be expressed in 
terms of correlation functions of a lattice model which is 
given by the action $\hat\1v$ and whose corresponding 
field configurations are functions from the faces of cubes into the 
interval $[0,1]$. Hence some properties of the 
non-commutative lattice field theory models  
$\Lam_n$ can be investigated by studying commutative lattice models.
This point of view may be helpful in order to construct 
non-ultra local models.

\paragraph{\it On the regularity condition WE3:}
However, the above discussion is not concerned with the the regularity 
condition {\em WE3}. By using a slightly different construction for the 
continuum C*-algebras we show in Section \ref{s3} how 
from a given family of invariant and reflexion positive states 
$\eta=(\eta_n)_{n\in\7N}$ continuum limits can be constructed 
which fulfill all the axioms of a weak euclidean statistical mechanics.
\abs

\paragraph{\it Conclusion and outlook:}
We close our paper by the Section \ref{s9} 
{\em conclusion and outlook}. 

\section{Continuum limits for lattice field theory models}
Within this section we develop a concept of {\em continuum limit}
which can be applied to a large class of lattice field theory 
models. In Section \ref{su1}, we introduce notation and conventions
which we are going to use.

A general and model independent notion of 
block spin transformations  is given in Section \ref{su2}.
Although the construction of C*-inductive limits is 
standard and can be found in many text books, we present a 
version of this procedure in Section \ref{su3}. One reason is 
to keep the paper as self contained as possible. 
Furthermore, 
the notations and definitions which we introduce 
in Section \ref{su3}, are used later to perform a procedure 
which is slightly different from taking the C*-inductive limit
of a net of C*-algebras. 

Finally, we present in Section \ref{su4}
a general concept for continuum limits of lattice models. 

\label{s1}
\subsection{Notation and conventions}
\label{su1}
We consider a C*-algebra
$A$ and for a given cutoff $n\in\7Z^2$ we introduce
the C*-algebra
\beqa
\6A_{(n,A)}&:=&\bigotimes_{\Delta\in \Sgm_d(n)} \{\Delta\}\times A \ \ .
\eeqa
To each subset $\9U\subset [-b^{n_1},b^{n_1}]^d$ a subalgebra
\beqa
\6A_{(n,A)}(\9U)&:=&\bigotimes_{\Delta\in \Sgm_d(n,\9U)} \{\Delta\}\times A
\eeqa
can be assigned in a natural manner. The set $\Sgm_d(n,\9U)$ is 
defined as follows: We identify the set 
$\Sgm^o_0(n):=b^{-n^0}\7Z^d\cap [-b^{n^1},b^{n^1}]^d$ with a subset 
of the torus $\Sgm_0(n)$. The $q$-cubes $\Sgm^o_q(n)$ 
in $b^{-n^0}\7Z^d\cap [-b^{n^1},b^{n^1}]^d$ can also be identified with 
$q$-cubes in $\Sgm_q(n)$. The set $\Sgm_d(n,\9U)$ consists of 
all cubes $\Delta$ in $\Sgm^o_d(n)$ with $\Delta\subset\9U$. 

The group $b^{-n^0}\7Z^d\subset\8E(d)$  acts by automorphisms
covariantly on the algebra $\6A_n(A)$. In other words, 
there exists a group homomorphism 
\beqa
\be_n\in\8{Hom}(b^{-n^0}\7Z,\8{Aut}\6A_{(n,A)})
\eeqa
such that for each $g\in b^{-n^0}\7Z^d $ the equation 
\beqa
\be_{(n,g)}\6A_{(n,A)}(\9U)&=&\6A_{(n,A)}(g\9U)
\eeqa
holds. The automorphism $\be_{(n,g)}$ is simply given by
\beqa
\be_{(n,g)}\biggl[\bigotimes_{\Delta\in\Sgm_d(n)}\{\Delta,a(\Delta)\}\biggr]
&:=&
\bigotimes_{\Delta\in\Sgm_d(n)}\{\Delta,a(g^{-1}\Delta)\} \ \ .
\eeqa
There is one important automorphism which corresponds to the 
euclidean time reflexion. 
\beqa
x\rMapsto \te_\mu(x)&=& x-2(e_\mu\cdot x) \ e_\mu
\eeqa
where $y\cdot x$ is the canonical scalar product in $\7R^d$ and 
$e_\mu\in\7R^d$ is the unit vector with components 
$(e_\mu)_\nu=\delta_{\mu\nu}$.
For each $\mu=1,\cdots , d$ we consider the anti-linear involution
\beqa
j_{(n,\mu)}:\6A_{(n,A)}\rTo  \6A_{(n,A)}
\eeqa
which is given by 
\beqa
j_{(n,\mu)}\biggl[\bigotimes_{\Delta\in\Sgm_*(n)}\{\Delta,a(\Delta)\}\biggr]
&:=&
\bigotimes_{\Delta\in\Sgm_d(n)}\{\Delta,a(\te_\mu\Delta)^*\} \ \ .
\eeqa
Since $b$ is odd, the set of $d$-cubes can be decomposed into 
a union of three disjoint set 
\beqa
\Sgm_d(n)&=&\Sgm_d(n,\mu,0)\cup\Sgm_d(n,\mu,+)\cup\Sgm_d(n,\mu,-)
\eeqa
where $\Sgm_d(n,\mu,0)$ is a layer of $\te_k$-invariant $d$-cubes
and $\Sgm_d(n,\mu,+)$ is mapped onto $\Sgm_d(n,\mu,-)$ via $\te_k$.
Therefore, operators of the form 
\beqa
a&=&\bigotimes_{\Delta\in\Sgm_d(n,\mu,0)}\{\Delta,a(\te_k\Delta)\}
\eeqa
with $a(\Delta)^*=a(\Delta)$ are $j_{(n,k)}$-invariant. 
The algebra $\6A_{(n,A)}$ can be written as a tensor product
\beqa
\6A_{(n,A)}&=&\6A_{(n,A)}(\mu,0)\otimes \6A_{(n,A)}(\mu,+)\otimes \6A_{(n,A)}(\mu,-)
\eeqa
where $\6A_{(n,A)}(\mu,0)$ is stable under $j_{(n,\mu)}$ and 
$\6A_{(n,A)}(\mu,+)$ is mapped onto $\6A_{(n,A)}(\mu,-)$ via $j_{(n,\mu)}$.
 
\subsection{Block spin transformations: The general setup}
\label{su2}
In our context, block spin
identify operators in $\6A_{(n,A)}$ with operators 
contained in a algebra $\6A_{(n_1,A)}$ which 
corresponds to a finer lattice, i.e. 
$n\prec n_1$. Let us state a list of axioms which 
characterizes the notion of block spin transformations.

\bdef
A family of *-homomorphisms
\beqa
\iota&=&\{\iota_{(n_1,n_0)}\in\8{Hom}(\6A_{(n_0,A)},\6A_{(n_1,A)})|
n_0\prec n_1\}
\eeqa
is called a family of {\em block spin transformations} 
if it fulfills the following conditions:
\bdes

\itno 1 
Cosheaf condition:
For each $n_0\prec n_1\prec n_2$:
\beqa
\iota_{(n_2,n_1)}\circ\iota_{(n_1,n_0)}&=&\iota_{(n_2,n_0)} \ \ .
\eeqa
\itno 2
Locality:
For each $n_0\prec n_1$ and for each $\9U\subset [-b^{n^1_0},b^{n^1_0}]^d$:
\beqa
\iota_{(n_1,n_0)}\6A_{(n_0,A)}(\9U)&\subset& \6A_{(n_1,A)}(\9U) \ \ .
\eeqa

\itno 3
Translation covariance:
Let $\9U_0,\9U_1\subset [-b^{n^1_0},b^{n^1_0}]^d$ such that 
$\9U_0\cup g\9U_0\subset \9U_1$ for some 
$g\in b^{-n^0_0}\7Z^d\subset b^{-n^0_1}\7Z^d $:
\beqa
\iota_{(n_1,n_0)}\beta_{(n_0,g)}a &=&
\beta_{(n_1,g)}\iota_{(n_1,n_0)}a
\eeqa
for each $a\in \6A_{(n_0,A)}(\9U_0)$.

\edes
\eef

\subsection{C*-inductive limits revisited}
\label{su3}
For a given family of block spin  transformations 
$\iota$ we construct the C*-inductive limit
$\6A_{(\iota,A)}$ of the net $n\mapsto \6A_{(n,A)}$.
In order to carry through our subsequent analysis, 
we briefly describe the construction of $\6A_{(\iota,A)}$.

\paragraph{\it Step I.} 
Let $\9C_\8b(\7Z^2,\6A_A)$ 
be the C*-algebra which is generated by bounded sections
in the bundle $\6A_A:n\mapsto\6A_{(n,A)}$.
We consider the closed two-sided ideal $\9C_0(\7Z^2,\6A_A)$
in $\9C_\8b(\7Z^2,\6A_A)$, which is generated by sections 
$a:n\mapsto a_n$ for which the limit 
$\lim_{n\to\infty} \|a_n\|=0$ vanishes. We build the quotient C*-algebra 
\beqa
\9C_\8a(\7Z^2,\6A_A)&:=&\9C_\8b(\7Z^2,\6A_A)/\9C_0(\7Z^2,\6A_A) \ \ .
\eeqa
In the following, $\1p$ denotes the corresponding canonical projection onto 
the quotient. 

\paragraph{\it Step II.} 
For a given family of block spin  transformations 
$\iota$, we denote by $\6A^o_{(\iota,A)}$ the 
C*-subalgebra in  $\9C_\8b(\7Z^2,\6A_A)$ which is generated by sections 
$a:n\mapsto a_n$ for which there exists 
$n_0\in\7Z^2$ and there exists $a_0\in \6A_{(n_0,A)}$ such that  
\beqa
a_n&=&\iota_{(n,n_0)}a_0
\eeqa
for each $n_0\prec n$. The C*-inductive limit of the pair $(\iota,A)$ now  
is given by 
\beqa
\6A_{(\iota,A)}&:=&\1p[\6A^o_{(\iota,A)}] \ \subset \ \9C_\8a(\7Z^2,\6A_A) \ \ .
\eeqa
For each $n\in\7Z^2$ we obtain a *-homomorphism 
$\iota_n:\6A_{(n,A)}\to \6A_{(\iota,A)}$ which identifies 
$\6A_{(n,A)}$ with a subalgebra in $ \6A_{(\iota,A)}$. It is given 
by the prescription
\beqa
\iota_n a&:=&\1p[ n_1\mapsto \iota_{(n_1,n)}a ] 
\eeqa
where the section $a_o=[n_1\mapsto \iota_{(n_1,n)}a]$ is any representative 
such that $a_o(n_1)=\iota_{(n_1,n)}a$ for each $n\prec n_1$.
It is obvious that the relation
$\iota_n\circ\iota_{(n,n_0)}=\iota_{n_0}$
holds for $n_0\prec n$. 

The C*-algebra $\6A_{(\iota,A)}$ can be regarded as the continuum C*-algebra
and it contains observables which correspond to observations at 
the full energy momentum range, whereas  
The C*-subalgebras $\iota_n(\6A_{(n,A)})$ contain only observables which 
correspond to observations for the energy momentum range  
$[b^{-n^1},b^{n^0}]$. 

We consider the dense subgroup  $\7Q_b^d:=\cup_n b^{-n}\7Z^d$ of the 
translation group $\7R^d$. There exists a group homomorphism 
\beqa
\beta_\iota\in\8{Hom}(\7Q_b^d,\8{Aut}\6A_{(\iota,A)})
\eeqa
which acts covariantly on $\6A_{(\iota,A)}$. For $g\in \7Q_b^d$ we define
\beqa
\beta_{(\iota,g)}\1p[ n\mapsto \iota_{(n,n_0)}a ]
&:=&
\1p[ n\mapsto \beta_{(n,g)}\iota_{(n,n_0)}a ]
\eeqa
with $g\in b^{-l}\7Z^d$ and $n^0>l$. Let $\6A_{(\iota,A)}(\9U)$ be the 
C*-subalgebra which is generated by local operators 
in $\iota_n[\6A_{(n,A)}(\9U)]$ for some $n\in\7Z^2$. Then we conclude 
from the construction of $\be_\iota$:
\beqa
\beta_{(\iota,g)}\6A_{(\iota,A)}(\9U)&=&\6A_{(\iota,A)}(g\9U) \ \ .
\eeqa
Thus the prescription
\beqa
\ul{\6A}_{(\iota,A)}:\9U\rMapsto \6A_{(\iota,A)}(\9U)
\eeqa
is a (weak) euclidean net of C*-algebras which is translationally covariant 
with respect to the group $\7Q_b^d$.

\subsection{On a general concept for continuum limits for lattice models}
\label{su4}
For each cutoff $n\in\7Z^2$ we select a class of appropriate states 
on $\6A_{(n,A)}$. We denote by $\6S_{(n,A)}$ the set of all states 
$\eta\in\6S(\6A_{(n,A)})$ which satisfy the assumptions:

\paragraph{\it Invariance:} 
For each $g\in b^{-n^0}\7Z$:
\beqa
\eta\circ\be_{(n,g)}&=&\eta \ \ .
\eeqa

\paragraph{\it Reflexion positivity:} 
The sesqui-linear form 
\beqa
a\otimes b\rMapsto \<\eta, j_{(n,\mu)}(a) b \>
\eeqa
is positive semi-definite on $\6A_{(n,A)}(\mu,+)$ for each $\mu=1,\cdots, d$.
\abs

There are also anti-linear involutions $j_{(\iota,\mu)}$ acting on the 
C*-inductive limit $\6A_{(\iota,A)}$ according to the prescription:
\beqa
j_{(\iota,\mu)}\1p[ n\mapsto \iota_{(n,n_0)}a ]
&:=&
\1p[ n\mapsto \iota_{(n,n_0)}j_{(n_0,\mu)}a ]
\vs\vs
&=&
\1p[ n\mapsto j_{(n,\mu)}\iota_{(n,n_0)}a ] \ \ .
\eeqa
Analogously to the definition, given above, we introduce the 
space $\6S_{(\iota,A)}$ of $\7Q^d_b$-invariant and 
reflexion positive functionals on $\6A_{(\iota,A)}$. 

Let $\Gam(\7Z^2,\6S_A)$ be the convex set of sections 
\beqa
\eta:\7Z^2\ni n\rMapsto \eta_n\in \6S_{(n,A)}
\eeqa
We identify $\6S_{(\iota,A)}$ with the corresponding subset 
in $\Gam(\7Z^2,\6S_A)$ by identifying  $\om\in\6S_{(\iota,A)}$:
with the section 
\beqa
\om:n\rMapsto\om_n:=\om\circ\iota_n \ \ .
\eeqa
For simplicity, we do not distinguish
the state $\om\in\6S_{(\iota,A)}$ and the corresponding section 
within our notation.

\bpro\label{contlim}
There is a canonical surjective convex-linear map 
\beqa
\1E:\6S[\9C_\8a(\7Z^2,\7C)]\otimes\Gam(\7Z^2,\6S_A)
\rTo \6S_{(\iota,A)} \ \ .
\eeqa
\epro
\bpr
For a state $\xi\in \6S[\9C_\8a(\7Z^2,\7C)]$ and a section 
$\eta\in\Gam(\7Z^2,\6S_A)$ we define a new section 
$\1E[\xi\otimes\eta]$ by 
\beqa
\<\1E[\xi\otimes\eta]_n,\1p[n\mapsto \iota_{(n,n_0)}a]\>
&:=&
\<\xi,\1p[ \ n\mapsto \ \<\eta_n,\iota_{(n,n_0)}a\> \ ]\> \ \ .
\eeqa
It is obvious that $\1E$ is convex linear and that 
$\1E[\xi\otimes\eta]$ fulfills the consistency condition
\beqa
\1E[\xi\otimes\eta]_n\circ\iota_{(n,n_0)}&=&\1E[\xi\otimes\eta]_{n_0} \ \ .
\eeqa 
Let $\om\in \6S_{(\iota,A)}$ be given, then we obtain 
by a straight forward computation 
\beqa
\1E[\xi\otimes\om ]&=&\om
\eeqa
for each $\xi\in \6S[\9C_\8a(\7Z^2,\7C)]$. Thus $\1E$ is surjective.
Finally, the invariance and the reflexion positivity follow 
directly from the construction of $\1E$.
\epr

\paragraph{Remark:}

\bdes
\itno 1
For a given family 
of lattice field theory models 
\beqa
\Lam_n:=
(\ul{\6A}_{(n,A|\eta)},\be_n,\eta_n,\7R^d,b^{-n^0}\7Z^d,\9K^d_n )_{n\in\7N} 
\eeqa
we introduce the set of continuum limits by 
\beqa
\6S_{(\iota,A)}[\eta]
&:=&\biggl\{\1E[\xi\otimes \eta ] \ \biggm | 
\ \xi\in \6S[\9C_\8a(\7Z^2,\7C)] \biggr\}
\ \subset \ \6S_{(\iota,A)} \ \ .
\eeqa

\itno 2
Proposition \ref{contlim} suggests a guideline how to 
construct continuum limits from an appropriate family 
$(\Lam_n)_{n\in\7N}$ 
of lattice field theory models. 
For each continuum limit $\varphi\in\6S_{(\iota,A)}[\eta]$ 
the statistical mechanics 
\beqa
(\ul{\6A}_{(\iota,A|\varphi)},\be_\iota,\varphi,\7R^d,\7Q_b^d,\9K^d) 
\eeqa
fulfill the axioms of a weak euclidean statistical mechanics
except the continuity requirement {\it WE3}. 
Hence we deal with a well posed problem, namely 
to analyze the properties of the states contained in 
$\6S_{(\iota,A)}[\eta]$ with respect to the 
properties of the section $\eta$. 
\edes

\section{Actions, effective actions, and continuum limits}   
\label{s2}
This section is destined to introduce the notions 
action and effective action within the C*-algebraic setup.
Section \ref{su5} is concerned with the problem of constructing 
from a given a weak euclidean
statistical mechanics a new model by means of perturbations.
For this purpose, we introduce the concept of action 
and effective action.

In particular, we study perturbations of ultra local models. 
We present in Section \ref{su6} a simple example for a family 
of block spin transformation which allows to compute some 
useful expressions quite explicitly.  

In Section \ref{su7}, 
we show that, for a given lattice, there is a large variety of 
reflexion positive invariant states. 

A criterion for the existence of an effective action for 
continuum limits is formulated in Section \ref{su8}. 

\subsection{Effective actions and continuum limits}
\label{su5}
To begin with, we consider for a 
weak euclidean statistical mechanics 
\beqa
\Lam&=&(\ul{\6A}_{(\iota,A)},\be_\iota,\om,\7R^d,\7Q_b^d,\9K^d)
\eeqa
where $A$ is a C*-algebra and $\iota$ a family of block spin 
transformations. The corresponding subsystems 
with respect to a finite lattice are 
\beqa
\Lam_n&=&(\ul{\6A}_{(n,A)},\be_n,\om_n,\7R^d,b^{-n^0}\7Z^d,\9K_n^d) 
\ \ \prec \ \
\Lam 
\eeqa
with $\om_n=\om\circ\iota_n$.
We are now interested in the problem of {\em deforming} the theory $\Lam$ in
such a way that one obtains a new one. 

We assume that the C*-algebras 
$\6A_{(n,A)}$, $\6A_{(\iota,A)}$, are von Neumann algebras, 
acting on separable Hilbert spaces 
$\2H_n$, $\2H$, 
and the states $\om_n=\<\Om_n,(\cdot)\Om_n\>$,
$\om=\<\Om,(\cdot)\Om\>$, are induced by 
a vector $\Om_n\in\2H_n$, $\Om\in\2H$, respectively, which are 
cyclic and separating for the corresponding algebras.
In order to study perturbations 
of the state $\om,\iota$, we introduce the notion  
{\em action}.
\bdef
We denote by $\9B_{(\om)}(\7Z^2,\6A_{A}')$ the set of all sections
\beqa
\1v:\7Z^2\ni n\rMapsto \1v_n\in\6A_{(n,A)}'
\eeqa
for which $\1v_n>0$ for each $n\in\7Z^2$ and for which the state
$\eta_{(\om,\1v,n)}$, given by 
\beqa
\<\eta_{(\om,\1v,n)},a\>&:=& \1z_{(\om,\1v,n)}^{-1}\< \om_n,\1v_n \ a\> \ \ ,
\eeqa
is contained in $\6S_{(n,A)}$. The section $\1v$ is called 
{\em action} and $\1z_{(\om,\1v,n)}=\<\om_n,\1v_n\>$
is called the {\em partition function} with respect to the triple
$(\om,\1v,n)$.
\eef

In oder to introduce the notion effective action, we consider  
for each $n_0\prec n$  the normal conditional expectation 
\beqa
\1e_{(\om,n_0,n)}:\6A_{(n,A)}'\rTo  \6A_{(n_0,A)}'
\eeqa
which is determined by the condition 
\beqa
\<\om_n,b \ \iota_{(n,n_0)}(a)\> 
&=&\<\om_{n_0},\1e_{(\om,n_0,n)}(b)a\> 
\eeqa
for each $b\in\6A_{(n,A)}'$ and for each $a\in\6A_{(n_0,A)}$.  

For a given action $\1v$  and for 
$k\in\7N^2$ we get a further action by 
\beqa
\1e^{(k)}_{(\om )}(\1v)_n&:=& \1e_{(\om,n,n+k)}(\1v_{n+k})  
\eeqa
and $\1e^{(k)}_{(\om )}(\1v)$ is called 
the {\em effective action} with respect to $k$ and $\1v$.
Note that $\1e^{(k)}_{(\om)}$ is a convex linear map from 
$\9B_{(\om)}(\7Z^2,\6A_{A}')$ into  $\9B_{(\om)}(\7Z^2,\6A_{A}')$.
To carry through our subsequent analysis we select an appropriate class
of actions in $\9B_{(\om)}(\7Z^2,\6A_A')$.

\bdef
We denote by $\Gam^o_{(\om)}(\7Z^2,\6A_{A}')$ the linear space of sections 
\beqa
\1f:n\rMapsto \1f_n\in \6A_{(n,A)}'
\eeqa
for which the semi-norms 
\beqa
[[\1f]]_{(\om,n)}&=&
\sup_{k\in\7N^2}\|\1e_{(\om,n,n+k)}(\1f_{n+k})\|
\eeqa
are finite for each $n\in\7Z^2$. The closure with respect to the 
corresponding Fr\'echet topology is denoted by   
$\Gam_{(\om )}(\7Z^2,\6A_{A}')$. Furthermore, we introduce the convex subset  
\beqa
\9A_{(\om )}(\7Z^2,\6A_{A}')&:=&
\8{cls}[\Gam^o_{(\om )}(\7Z^2,\6A_{A}')\cap\9B_{(\om)}(\7Z^2,\6A_{A}')] \ \ .
\eeqa
\eef
\abs

\paragraph{Remark:}
\bdes
\itno 1
Note that the norms may increase with $n$, i.e.
\beqa
[[\1f]]_{(\om,n)}&\leq&[[\1f]]_{(\om,n_1)}
\eeqa  
for $n\prec n_1$. 
\itno 2
The maps $\1e^{(k)}_{(\om )}$ are continuous maps.
For a fixed cut-off $n\in\7Z^2$ the operation of 
$\1e^{(k)}_{(\om )}$ corresponds 
to a substitution by the underlying lattice theory on 
$\Sgm_d(n)$ by a lattice theory, also defined on 
$\Sgm_d(n)$, arising from a lattice theory 
on $\Sgm_d(n+k)$ by integrating out the corresponding high energy 
degrees of freedom. An action $\1v$ which is stable under 
$\1e^{(k)}_{(\om)}$ for every $k\in\7N^2$ can be interpreted as 
a {\em continuum limit}. As we shall see below, 
this can be justified by the fact that 
then the section $\eta_{(\om,\1v)}$ is contained in $\6S_{(\iota,A)}$, i.e. 
\beqa
\eta_{(\om,\1v,n)}\circ\iota_{(n,n_0)}&=&\eta_{(\om,\1v,n_0)} \ \ .
\eeqa
\edes

In order to point out the structure of the 
space $\Gam_{(\om )}(\7Z^2,\6A_{A}')$ and the cone 
$\9A_{(\om)}(\7Z^2,\6A_{A}')$, we summarize some facts in the proposition 
below.
\bpro
\bdes

\itno {i}
For each $\7Z^2$-invariant state $\xi\in \6S[\9C_\8a(\7Z^2,\7C)]$ 
there exists a continuous linear map 
\beqa
\1e_{(\om,\xi)}:\Gam_{(\om)}(\7Z^2, \6A_{A}')\rTo
\Gam_{(\om )}(\7Z^2,\6A_{A}')
\eeqa
such that for each 
$k\in\7N^2$ the following holds true:
\beqa
\1e^{(k)}_{(\om )}\circ \1e_{(\om ,\xi)}&=&\1e_{(\om ,\xi)} \ \ .
\eeqa

\itno {ii}
For each $\1v\in \9A_{(\om)}(\7Z^2,\6A_{A}')$ the state 
$\eta_{(\om,\1e_{(\om,\xi)}(\1v))}$ is contained 
in $\6S_{(\iota,A)}$.

\edes
\epro
\bpr
\bdes
\itno {i}
For each $\1f\in\Gam_{(\om )}(\7Z^2,\6A_{A}')$ and for each $n\in\7Z^2$
we obtain a bounded family of operators 
$(\1e_{(\om ,n,n+k)}(\1f_{n+k}))_{k\in\7N^2}$ in $\6A_{(n,A)}'$ since 
the semi-norm $[[\1f]]_{(\om ,n)}$ is finite. 
For any bounded family $(\1w_k)_{k\in\7N^2}$ and for a given state 
$\xi\in \6S[\9C_\8a(\7Z^2,\7C)]$ we define a continuous linear 
map on the pre-dual $(\6A_{(n,A)}')_*$ by 
\beqa
\1w_\xi:\varphi\rMapsto \<\xi,\1p[k\mapsto \<\varphi,\1w_k\>]\> 
\eeqa
and hence $\1w_\xi\in\6A_{(n,A)}'$. We define the map  $\1e_{(\om ,\xi)}$
according to 
\beqa
\1e_{(\om,\xi)}(\1f)_n&:=& \1e_{(\om,n,\xi)}(\1f)
\eeqa
where $\1e_{(\om ,n,\xi)}(\1f)$ is given by 
\beqa
\1e_{(\om,n,\xi)}(\1f):\varphi\rMapsto 
\<\xi,\1p[k\mapsto \<\varphi,\1e_{(\om ,n,n+k)}(\1f_{n+k})\>]\> \ \ .
\eeqa
We have for each $a\in\6A_{(n,A)}$:
\beqa
&&\<\om_n,\1e_{(\om,n,n+k_0)}\1e_{(\om,n+k_0,\xi)}(\1f)a\>
\vs\vs
&=&
\<\om_{n+k_0},\1e_{(\om,n+k_0,\xi)}(\1f)\iota_{(n+k_0,n)}(a)\>
\vs\vs
&=&
\<\xi,\1p[k\mapsto \<\om_{n+k+k_0},\1f_{n+k+k_0}
\iota_{(n+k+k_0,n)}(a)\>]\>
\vs\vs
&=&
\<\xi,\1p[k\mapsto \<\om_{n+k},\1f_{n+k} 
\iota_{(n+k,n)}(a)\>]\>
\vs\vs
&=&
\<\om_n,\1e_{(\om,n,\xi)}(\1f)a\> 
\eeqa
which yields 
\beqa
\1e_{(\om,n,n+k_0)}\1e_{(\om ,n+k_0,\xi)}(\1f)&=&
\1e_{(\om ,n,\xi)}(\1f) \ \ .
\eeqa
Finally we conclude  
\beqa
[[\1e_{(\om ,\xi)}(\1f)]]_{(\om ,n)} &=&
\sup_{k\in\7N^2}\|\1e_{(\om,n,n+k)}\1e_{(\om ,n+k,\xi)}(\1f)\| 
\vs\vs
&=&
\sup_{k\in\7N^2}\|\1e_{(\om ,n,n+k)}\1e_{(\om ,n+k,\xi)}(\1f)\| 
\vs\vs
&=&
\|\1e_{(\om,n,\xi)}(\1f)\| \ \leq \ [[\1f]]_{(\om ,n)}
\eeqa
which proves {\it (i)}.

\itno {ii}
Let $\1v\in\Gam^o_\om(\7Z^2,\6A_{A}')\cap\9B_{(\om)}(\7Z^2,\6A_{A}')$ 
be an action. For each $n\in\7Z^2$ the state 
$\eta_{(\om,\1v,n)}$ is reflexion positive and $b^{-n^0}\7Z^d$-invariant.
For each $n\in\7Z^2$ and for each $a\in\6A_{(n,A)}$ we compute
for an action $\1v'$ which is stable under $\1e^{(k)}_{(\om)}$
for each $k\in\7N^2$: $\1e^{(k)}_{(\om)}(\1v')=\1v'$:
\beqa
\<\om_{n+k},\1v'_{n+k}\iota_{(n+k,n)}a \>
&=&
\<\om_n,\1e_{(\om,n,n+k)}(\1v'_{n+k}) a \>
\vs\vs
&=&
\<\om_n,\1v'_n a \>
\eeqa
which yields for $a=1$:
\beqa
\1z_{(\om,\1v',n+k)}
&=&
\<\om_n,\1e_{(\om,n,n+k)}(\1v'_{n+k}) a \>
\vs\vs
&=&
\1z_{(\om,\1v',n)}
\eeqa
and therefore 
\beqa
\1E[\xi'\otimes\eta_{(\om,\1v')}]&=&\eta_{(\om,\1v')}
\eeqa
for each $\xi'$, and thus we conclude for $\1v'=\1e_{(\om,\xi)}(\1v)$:
\beqa
\1E[\xi\otimes\eta_{(\om,\1v)}]&=&
\1E[\xi'\otimes\eta_{(\om,\1e_{(\om,\xi)}(\1v))}]
\vs\vs
&=&
\eta_{(\om,\1e_{(\om,\xi)}(\1v))}
\eeqa
which implies $\eta_{(\om,\1e_{(\om,\xi)}(\1v))}\in\6S_{(\iota,A)}$.

Let $(\1v_i)_{i\in I}$ be a net in 
$\Gam^o_\om(\7Z^2,\6A_A')\cap\9A(\7Z^2,\6A_A')$ 
which converges to $\1v$ in $\9A_{(\om)}(\7Z^2,\6A_A')$.
For each $n\in\7Z^2$ and for each $a\in\6A_{(n,A)}$ the map
\beqa
\1T_{(a,\om,n)}:\1f\rMapsto \<\om_n,\1e_{(\om,n,\xi)}(\1f) a \>
\eeqa
is a {\em continuous} linear functional on $\Gam_{(\om)}(\7Z^2,\6A_{A}')$
which follows directly from the estimate
\beqa
|\<\om_n,\1e_{(\om,n,\xi)}(\1f) a \>| &\leq&
\|\1e_{(\om,n,\xi)}(\1f)\| \ \|a\|
\vs\vs
&\leq&
[[\1f]]_{(\om,n)} \ \|a\| \ \ .
\eeqa
Therefore we have
\beqa
\<\om_n,\1e_{(\om,n,\xi)}(\1v) \be_{(n,g)}a \>
&=&
\<\om_n,\1e_{(\om,n,\xi)}(\lim_{i\in I}\1v_i) \be_{(n,g)}a \>
\vs\vs
&=&
\lim_{i\in I}\<\om_n,\1e_{(\om,n,\xi)}(\1v_i) \be_{(n,g)}a \>
\vs\vs
&=&
\lim_{i\in I}\<\om_n,\1e_{(\om,n,\xi)}(\1v_i)a \>
\vs\vs
&=&\<\om_n,\1e_{(\om,n,\xi)}(\1v) a \>
\eeqa
which proves the $\7Q^d_b$-invariance of $\eta_{(\om,\1e_{(\om,\xi)}(\1v))}$.
The reflexion positivity follows by an analogous argument.
\edes
\epr
\abs

We formulate one important consequence of the proposition above 
by the following corollary:

\bcor\label{fiequ} 
For each action $\1v\in \9A_{(\om)}(\7Z^2,\6A_A')$ and for each 
continuum limit $\eta\in\6S_{(\iota,A)}[\eta_{(\om,\1v)}]$
the states 
\beqa
\eta\circ\iota_n \in \6S_{(n,A)}\cap (\6A_{(n,A)})_* 
\eeqa
are normal for each $n\in\7Z^2$.
\ecor
\bpr
For each operator $a\in\6A_{(n,A)}$ we have for a continuum limit 
$\eta=\1E[\xi\otimes \eta_{(\om,\1v)}]$
\beqa
\<\1E[\xi\otimes \eta_{(\om,\1v)}],\iota_n(a)\>
&=&\<\eta_{(\om,n)},\1e_{(\om,n,\xi)}(\1v) a \>
\eeqa
which proves the normality.
\epr

\paragraph{Remark:}
\bdes
\itno 1
A given action $\1v\in \9A_{(\om)}(\7Z^2,\6A_A')$ 
can be used to {\em deform} the given theory $\Lam$, namely 
for each  $\7Z^2$-invariant state $\xi\in \6S[\9C_\8a(\7Z^2,\7C)]$ we obtain  
a new theory 
\beqa
\Lam^{(\xi,\1v)}
&:=&(\ul{\6A}_{(\iota,A|\eta)},\be_\iota,\eta,\7R^d,\7Q_b^d,\9K^d) 
\eeqa
with $\eta=\eta_{(\om,\1e_{(\om,\xi)}(\1v))}$
and the net $\ul{\6A}_{(\iota,A|\eta)}$ is given by
\beqa
\ul{\6A}_{(\iota,A|\eta)}:\9U\rMapsto\6A_{(\iota,A)}(\9U)/\6J_{(\iota,\eta)} \ \ .
\eeqa
 
\itno 2
If $\6A_{(\iota,A)}$ is a factor of type III, then the theories 
$\Lam^{(\xi,\1v)}$ and $\Lam$ are inequivalent if and only if 
$\eta$ is not normal on $\6A_{(\iota,A)}$. 

\itno 3
If $\om$ is a gaussian state, the statement of Corollary \ref{fiequ} 
can be regarded as a weaken version of the {\em local Fock 
property} \cite{GlJa1,Schra1}. Whereas the local Fock property states that 
the restriction of the deformed state $\eta$  is normal 
on each local algebra,  Corollary \ref{fiequ} states that 
one also have to restrict to operators which correspond to a
high energy momentum cut-off. 

\itno 4
We claim here that a necessary condition for $\1v$ such that the 
states $\om$ and $\eta$ are disjoint is that the supreme 
\beqa
\sup_{n\in\7Z^2}\|\1v_n\|_{\6A_{(n,A)}'}&=&\infty
\eeqa
is infinite.
\edes
\subsection{Block spin transformations: Concrete examples}
\label{su6}
Let $M$ be a von Neumann algebra acting on a Hilbert space $K$
and let $\Om$ be a cyclic and separating vector for $M$.
We consider the von Neumann algebra 
\beqa
\6A_{(n,M)}&:=&\ol{\bigotimes_{\Delta\in\Sgm_d(n)}}\{\Delta\}\times M  
\eeqa
acting on the Hilbert space
\beqa
\2H_{(n,K)}&:=&\bigotimes_{\Delta\in\Sgm_d(n)}\{\Delta\}\times K \ \ .
\eeqa
The vector 
\beqa
\Om_n&:=& \bigotimes_{\Delta\in\Sgm_d(n)}\{\Delta,\Om\}
\eeqa
is cyclic and separating for $\6A_{(n,M)}$. 

For each $n_0\prec n$ we define for each cube 
$\Delta_0\in\Sgm_d(n_0) $ the cube   
\beqa
*\Delta_0\subset\Delta_{(n,n_0|\Delta_0)}\subset\Delta_0
\eeqa
in $\Sgm_d(n)$ which is determined by the condition to contain the 
dual one cube $*\Delta_0$ of $\Delta_0$.
A faithful normal *-homomorphism from $\6A_{(n_0,M)}$ into 
$\6A_{(n,M)}$ is given by 
\beqa
\iota_{(n,n_0)}&:=&\bigotimes_{\Delta_0\in\Sgm_d(n_0)}
\biggl[ \11_{D(n,n_0|\Delta_0)}\otimes\{\Delta_{(n,n_0|\Delta_0)},\id\}\biggr]
\eeqa
with $\11_D:=\otimes_{\Delta\in D}\{\Delta,\11\}$ for a subset  
$D\subset\Sgm_d(n)$. Here the set $D(n,n_0|\Delta_0)$ 
of hypercubes is 
\beqa
D(n,n_0|\Delta_0)&:=&\{\Delta\subset\Delta_0|
\Delta\not= \Delta_{(n,n_0|\Delta_0)} \} \ \  .
\eeqa
The following proposition follows directly from the definitions, 
given above.

\bpro
The family $\iota=(\iota_{(n,n_0)})_{n_0\prec n}$ is a family of 
block spin transformations. 
\epro

For each $n\in\7Z^2$ we consider the normal state 
$\om_n:=\<\Om_n,(\cdot)\Om_n\>$.
One easily verifies that the section 
$\om:n\mapsto\om_n$ satisfies the 
consistency condition with respect to 
$\iota$, i.e. 
\beqa
\om_n\circ\iota_{(n,n_0)} &=&\om_{n_0}
\eeqa
and therefore $\om=\1E[\xi\otimes\om]$ is a state 
on the C*-inductive limit $\6A_{(\iota,M)}$, 
independent of the choice of $\xi$. This yields a 
statistical mechanics 
\beqa
\Lam&:=& (\ul{\6A}_{(\iota,M)},\be_\iota,\om,\7R^d,\7Q_b^d,\9K^d)
\eeqa
which fulfills the axioms {\it WEF1} and {\it WEF2}. 

For each pair $n_0\prec n$ there is a normal conditional expectation 
\beqa
&&\1e_{(\om,n_0,n)}
\vs\vs 
&:=&
\bigotimes_{\Delta_0\in\Sgm_d(n_0)}
\biggl[ 
\biggl[ \bigotimes_{\Delta\in D(n,n_0|\Delta_0)} \{\Delta,\<\Om,(\cdot)\Om\>\}
\biggr] 
\ \otimes \ 
\{\Delta_{(n,n_0|\Delta_0)},\id\}\biggr]
\eeqa
which maps $\6A_{(n,M)}'$ into $\6A_{(n_0,M)}'$ and one easily computes for 
$b\in\6A_{(n,M)}'$ and for $a\in\6A_{(n_0,M)}$:
\beqa
\<\om_n,b \ \iota_{(n,n_0)}(a)\> &=& \<\om_{n_0},\1e_{(\om,n_0,n)}(b) a\> \ \ .
\eeqa

\subsection{Construction of invariant reflexion positive states}
\label{su7}
We are now interested in the space 
$\9B_{(\om)}(\7Z^2,\6A_M')$ of actions in order to perform deformations 
of the theory $\Lam$ which we have introduced in the previous section. 

Let $a\in M$ be an operator and let $\Delta\in\Sgm_d(n)$ be a cube, 
then we write $\Phi_n(\Delta,a)$ 
for the corresponding element in $\6A_{(n,M)}$.
For each face $\Gam\in\Sgm_{d-1}(n)$ there are unique cubes 
$\Delta_0,\Delta_1\in\Sgm_d(n)$ such that $\Gam=\Delta_0\cap\Delta_1$.
We write: $\Phi_n(\Gam,a\otimes b)=\Phi_n(\Delta_0,a)\Phi_n(\Delta_1,b)$ for 
$a,b\in M$.  Let $w=(w_n)_{n\in\7N}\subset M'\otimes M'$ be a 
family of positive operators such that 
\beqa
[w_n\otimes\11,\11\otimes w_n]&=&0
\eeqa
for each $n\in\7Z^2$. Then we introduce for $n\in\7Z^2$ the positive operator 
\beqa
\1v[w]_n&:=&\prod_{\Gam\in\Sgm_{d-1}(n)}\Phi_n(\Gam,w_n) \in \6A_{(n,M)}' 
\eeqa 
and we obtain a section 
\beqa
\1v[w]\in\Gam(\7Z^2,\6A_{M'}) \ \ .
\eeqa

\bpro\label{proact}
Given a family $w=(w_n)_{n\in\7N}\subset M'\otimes M'$ of positive 
operators such that
\beqa
[\11\otimes w,w\otimes\11]&=&0 \ \ ,
\eeqa
then the section $\1v[w]$ is an action contained in 
$\9B_{(\om)}(\7Z^2,\6A_M')$. 
\epro
\bpr
For each $n\in\7Z^2$ it is obvious, that the state 
$\eta_{(\om,\1v[w],n)}$ is $b^{-n^0}\7Z^d$ invariant. 
Let $\Sgm_{d-1}(n,\mu,0)$ be the subset in 
$\Sgm_{d-1}(n)$ which consists of all faces $\Gam$ of cubes in 
$\Sgm_d(n,\mu,0)$ which intersect the hyperplane $\Sgm_{e_k}$.
We define the sets 
\beqa
\Sgm_{d-1}(n,\mu,\pm)&:=&\{\Gam\in\Sgm_{d-1}(n)\bs\Sgm_{d-1}(n,\mu,0)|
\Gam\subset\7R_\pm e_k+\Sgm_{e_k}\}
\eeqa
and $\1v[w]_n$ can be decomposed as follows:
\beqa
\1v[w]_n&:=&\1v[w,0]_n\1v[w,+]_n\1v[w,-]_n
\eeqa
where $\1v[w,\ell]_n$, $\ell=0,\pm$ is given by 
\beqa
\1v[w,\ell]_n&:=&\prod_{\Gam\in\Sgm_{d-1}(n,\mu,\ell)}\Phi_n(\Gam,w_n)  \ \ .
\eeqa 
The operators $\1v[w,\pm]_n$ are contained in 
$\6A_{(n,M')}(\mu,0)\olt\6A_{(n,M')}(\mu,+)$ and we conclude
for an operator $a\in\6A_{(n,M)}(\mu,0)\olt\6A_{(n,M)}(\mu,+)$ 
\beqa
\1v[w]_n \ j_{(n,k)}(a) \ a &=&
\1v[w,0]_n\ j_{(n,k)}(\1v[w,+]_n a) \ \1v[w,+]_n a 
\vs\vs
&=&
\1v[w,0]_n^{1/2}\ j_{(n,k)}(\1v[w,+]_n a) \ \1v[w,+]_n a \1v[w,0]_n^{1/2} 
\ \ .
\eeqa
where we have used the fact that 
\beqa
j_{(n,k)}(\1v[w,+]_n) &=& \1v[w,-]_n
\vs\vs
{[ \1v[w,0]_n , \1v[w,\pm]_n ]} &=& 0 \ \ .
\eeqa 
We put for $\ell=0,\pm$
\beqa
\Om_{(n,\ell)}&:=&\bigotimes_{\Delta\in\Sgm_d(n,\mu,\ell)}
\{\Delta,\Om\} 
\eeqa
and we consider the conditional expectation 
\beqa
&&\1E_{(\om,n,k)}
\vs\vs
&:=&
\<\Om_{(n,+)},(\cdot)\Om_{(n,+)}\> \otimes
\<\Om_{(n,-)},(\cdot)\Om_{(n,-)}\>
\otimes\biggl[\bigotimes_{\Delta\in\Sgm_d(n,\mu,0)}\{\Delta,\id\}\biggr] \ \ .
\eeqa
We compute for operators $a_\pm\in\6A_{(n,\6B(K))}(\mu,\pm)$
and $b_\pm\in\6A_{(n,\6B(K))}(\mu,0)$: 
\beqa
&&
\1E_{(\om,n,k)}((a_-\otimes b_-) (a_+\otimes b_+))
\vs\vs
&=& 
b_-b_+\<\Om_{(n,+)},a_+\Om_{(n,+)}\> \<\Om_{(n,-)},a_-\Om_{(n,-)}\> 
\vs\vs
&=&
b_-b_+\1E_{(\om,n,k)}(a_-)\1E_{(\om,n,k)}(a_+)
\vs\vs
&=&
\1E_{(\om,n,k)}(a_-\otimes b_-)\1E_{(\om,n,k)}(a_+\otimes b_+)
\eeqa
which implies 
\beqa
&&\1E_{(\om,n,k)}(\ \1v[w]_n \ j_{(n,k)}(a) \ a \ ) 
\vs\vs
&=&
\1v[w,0]_n^{1/2} \ \1E_{(\om,n,k)}(\ j_{(n,k)}(\1v[w,+]_n a) \ )
 \ \1E_{(\om,n,k)}(\1v[w,+]_n a) \ \1v[w,0]_n^{1/2}
\vs\vs
&=&
\1v[w,0]_n^{1/2} \ \1E_{(\om,n,k)}(\ \1v[w,+]_n a \ )^* \ 
\1E_{(\om,n,k)}(\1v[w,+]_n a) \ \1v[w,0]_n^{1/2}\ \ .
\eeqa
Here we have used the fact that $\1E_{(\om,n,k)}$ is 
invariant under the euclidean time reflexion $j_{(n,k)}$.
We conclude for $\Psi_n:=\1v[w,0]_n^{1/2}\Om_{(n,0)}$ 
\beqa
&&\<\eta_{(\om,n)},\1v[w]_n \ j_{(n,k)}(a) \ a\>
\vs\vs
&=&
\<\Om_{(n,0)}, \ \1E_{(\om,n,k)}(\ \1v[w]_n \ j_{(n,k)}(a) \ a \ )\Om_{(n,0)}\>
\vs\vs
&=&
\<\Psi_n, \1E_{(\om,n,k)}(\ \1v[w,+]_n a \ )^* \ 
\1E_{(\om,n,k)}(\1v[w,+]_n a)\Psi_n\>
\vs\vs
&\geq& 0
\eeqa
which proves the reflexion positivity.
\epr

\subsection{Multiplicative renormalization}
\label{su8}
The main problem which arises here is to check 
that the set $\9A_{(\om)}(\7Z^2,\6A_{M'})$ contains interesting 
elements. Let $\1v\in\9B_{(\om)}(\7Z^2,\6A_{M'})$ be any action. 
According to what we claim in Section \ref{su5},  one has to deal with 
the following behavior
for the partition function, provided one requires that 
$\|\1v\|:=\sup_{n\in\7Z^2}\|\1v_n\| <\infty$: 
\beqa
\lim_{n\in\7Z^2}\1z_{(\om,\1v,n)} &=& 0 \ \ ,
\eeqa
in order to obtain a deformed theory  $\Lam^{(\xi,\1v)}$ 
which is not equivalent to the underlying one.

Furthermore, one expects that for each $n\in\7Z^2$
\beqa
\lim_{k\in\7N^2}\1e_{(\om,n,n+k)}(\1v_{n+k})&=&0
\eeqa
which yields $\1e_{(\om,\xi)}(\1v)=0$. In order to get a non-trivial 
limit we replace $\1v$ by 
\beqa
\1r_\om\1v:n\mapsto\1z_{(\om,\1v,n)}^{-1}\1v_n
\eeqa
This implies for $\1r_\om\1v$
\beqa
\1z_{(\om,\1r_\om\1v,n)} &=& 1
\eeqa
for each $n\in\7Z^2$ and therefore 
\beqa
1&\leq&\sup_{k\in\7N^2}\|\1e_{(\om,n,n+k)}(\1r_\om\1v_{n+k})\| \ = \ 
[[\1r_\om\1v]]_{(\om,n)}
\eeqa
provided the right hand side is finite.
The semi-norms of the resulting fix-points $\1e_{(\om,\xi)}(\1r_\om\1v)$ 
are bounded from below by $1$. The operation $\1r_\om$
can be regarded as {\em multiplicative 
renormalization}. Therefore it is natural to call the condition 
$\1r_\om\1v\in \9A_{(\om)}(\7Z^2,\6A_{M'})$ {\em multiplicative 
renormalizability} of $\1v$. 

We first illustrate the notion multiplicative renormalization
by an ultra-local example. 
Let $\1v:n\mapsto\1v_n$ be a section of the form
\beqa
\1v_n&=&\prod_{\Delta\in\Sgm_d(n)}\Phi_n(\Delta,w_n)
\eeqa
with $w_n\in M'$ and $\|w_n\|=1$ for each $n\in\7Z^2$. Then we 
easily compute 
\beqa
\1z_{(\om,\1v,n)}&=&\<\Om,w_n\Om\>^{\tau(n)}
\vs\vs
\1e_{(\om,n,n+k)}(\1v_{n+k})&=&\prod_{\Delta\in\Sgm_d(n)}
\Phi_{n+k}(\Delta,w_{n+k})
\<\Om,w_{n+k}\Om\>^{\tau(n+k)-\tau(n)} \ \ .
\eeqa
For $\limsup\<\Om,w_n\Om\> < 1$ the partition function $\1z_{(\om,\1v,n+k)}$
vanishes for $k\to\infty$. On the other hand we have 
\beqa
\1z_{(\om,\1v,n+k)}^{-1}\|\1e_{(\om,n,n+k)}(\1v_{n+k})\|
&\leq& \<\Om,w_{n+k}\Om\>^{-\tau(n)} 
\eeqa
with $\tau(n):=b^{d(n^0+n^1)}$.
By choosing $w_n=w$ with $\<\Om,w\Om\>=\gam<1$, for instance, we conclude
\beqa
1\ \ < \ \ [[\1r_\om\1v]]_{(\om,n)}&\leq&\gam^{-\tau(n)} \ \ < \ \ \infty
\eeqa
and $\1v$ is multiplicatively renormalizable. 
An example for a multiplicatively non-renormalizable action  
can be obtained by choosing $(w_n)_{n\in\7Z^2}$ in such a way that 
$\lim_n\<\Om,w_n\Om\>=0$. 

From the physical point of view, perturbation 
of $\om$ by ultra local action 
is quite uninteresting since the corresponding theory 
in Minkowski space, provided it exists, is then nothing else but the
constant field. In the subsequent, we discuss conditions 
under which a non-ultra local action is multiplicatively
renormalizable.

Let $h\in\9C^\infty(\7Z^2\times [0,1],M')$ 
be an operator-valued function which is smooth in its second variable and 
for which  $[h(n,s_1),h(n,s_2)]=0$ for each $s_1,s_2\in[0,1]$
and for which $\|h(n,s)\|\leq 1$. We introduce the following 
numbers in $\7R_+\cup\{\infty\}$ associated with $h$:
\beqa
I_{(\om,n)}(h)&:=&\inf_{s_1\cdots s_{2d}}\<\Om, h(n,s_1)\cdots h(n,s_{2d})\Om\>
\vs\vs
S_{(\om,n)}(h)&:=&\sup_{s_1\cdots s_{2d}}\<\Om, h(n,s_1)\cdots h(n,s_{2d})\Om\>
\vs\vs
R_{(\om,n)}(h)&:=&
\sup_{k\in\7N^2}\biggl( { S_{(\om,n+k)}(h)\over 
I_{(\om,n+k)}(h) }\biggr)^{\tau(n+k)}S_{(\om,n+k)}(h)^{-\tau(n)}
\eeqa 
and we define an action by  
\beqa
\1v[h]_n &:=&\prod_{\Gam\in\Sgm_{d-1}(n)}\Phi_n\biggl( \Gam , 
\int_0^1 \8ds \ h(n,s)\otimes h(n,s) \biggr)
\vs\vs
&=&\int \prod_{\Gam\in\Sgm_{d-1}(n)}\8ds(\Gam)
\ \bigotimes_{\Delta\in\Sgm_d(n)} \biggl\{\Delta,\prod_{\Gam\in\pa\Delta}
h(n,s(\Gam))\biggr\} \ \ .
\eeqa
The proposition, given above states a sufficient condition
for $h$ such that $\1v[h]$ can multiplicatively by renormalized.

\bpro\label{themultren}
Let $h\in\9C^\infty(\7Z^2\times [0,1],M')$ be given. If 
$R_{(\om,n)}(h)<\infty$ is finite for every $n$,
then the action $\1v[h]$ is multiplicatively renormalizable, 
i.e. $\1r_\om\1v[h]\in\9A_{(\om)}(\7Z^2,\6A_{M'})$.
\epro
\bpr
Computing the partition function gives 
\beqa
\1z_{(\om,\1v[h],n)}
&=&
\int \prod_{\Gam\in\Sgm_{d-1}(n)}\8ds(\Gam)
\ \prod_{\Delta\in\Sgm_d(n)} 
\bla \om ,\prod_{\Gam\in\pa\Delta}h(n,s(\Gam)) \bra
\eeqa
and according to our assumption the partition function 
$\1z_{(\om,\1v[h],n)}$ satisfies the 
inequality
\beqa
I_{(\om,n)}(h)^{\tau(n)} \ \ \leq \ \ 
\1z_{(\om,\1v[h],n)} \ \ \leq \ \ S_{(\om,n)}(h)^{\tau(n)}  \ \ .
\eeqa
and we compute 
\beqa
\1e_{(\om,n,n+k)}(\1v[h]_{n+k})
&=&
\int \prod_{\Gam\in\Sgm_{d-1}(n+k)}\8ds(\Gam)
\bigotimes_{\Delta_0\in\Sgm_{d}(n)}\biggl[
\vs\vs
&\times&
\prod_{\Delta\in\ D(n,n+k|\Delta_0)} 
\bla \om ,\prod_{\Gam\in\pa\Delta}h(n+k,s(\Gam)) \bra
\vs\vs
&\times&
\biggl\{\Delta_0, \prod_{\Gam\in\pa\Delta_{(n,n+k|\Delta_0)}}h(n+k,s(\Gam))
\biggr\} \biggr]
\eeqa
which implies for the norm
\beqa
\|\1e_{(\om,n,n+k)}(\1v[h]_{n+k})\|&\leq&
S_{(\om,n+k)}(h)^{\tau(n+k)-\tau(n)} \ \ .
\eeqa
This yields
\beqa
&&\1z_{(\om,\1v[h],n+k)}^{-1} \|\1e_{(\om,n,n+k)}(\1v[h]_{n+k})\|
\vs\vs
&\leq&
I_{(\om,n+k)}(h)^{-\tau(n+k)}S_{(\om,n+k)}(h)^{\tau(n+k)-\tau(n)} 
\vs\vs
&\leq&
\biggl({S_{(\om,n+k)}(h)\over I_{(\om,n+k)}(h)}\biggr)^{\tau(n+k)}
S_{(\om,n+k)}(h)^{-\tau(n)}
\eeqa
and we obtain for the semi-norms the estimate
\beqa
[[ \ \1v[h] \ ]]_{(\om,n)} &\leq& R_{(\om,n)}(h) 
\eeqa
and the result follows.
\epr

\section{Weak euclidean field theory models}
\label{s3}
As already mentioned, the previous sections are not concerned with 
the regularity condition {\it WE3}. 
In Section \ref{su10} we 
present a procedure which, in comparison to building the 
C*-inductive limit, leads to  a euclidean net of C*-algebras
on which the full euclidean group acts by automorphisms. In particular
we show that the translations act norm continuously. 

Section \ref{su11} is concerned with states which fulfill 
all axioms for weak euclidean statistical mechanics. We show that {\em each} 
section of states $\eta\in \Gam(\7Z^2,\6S_A)$ can be associated with 
a family of weak euclidean statistical mechanics. 
 
\subsection{Construction of a weak euclidean net of C*-algebras}
\label{su10}
For a given C*-algebra $A$, we consider for each $n\in\7Z^2$ 
the tensor algebra $\6T_{(n,A)}:=T(S(\7R^d)\otimes\6A_{(n,A)})$
over the linear space $S(\7R^d)\otimes\6A_{(n,A)}$. For a region 
$\9U\subset \7R^d$ we denote by $\6T_{(n,A)}(\9U)$ 
the *-subalgebra in $\6T_{(n,A)}$ which is generated by 
operators $f\otimes a$ with $a\in\6A_{(n,A)}(\9U_0)$ and 
$\supp(f)+\9U_0\subset\9U$. 
For each $n\in\7Z^2$ there is a group homomorphism
\beqa
\tau_n\in\8{Hom}(\8E(d),\8{Aut}(\6T_{(n,A)}))
\eeqa
which is defined by 
\beqa
\tau_{(n,g)}(f\otimes a)&:=&(f\circ g^{-1})\otimes a \ \ .
\eeqa
It is obvious that the euclidean group $\8E(d)$ acts covariantly on 
the net $\ul{\6T}_{(n,A)}:\9U\mapsto\6T_{(n,A)}(\9U)$.

For each $n\in\7Z^2$ we now introduce a *-homomorphism $\Phi_n$ which maps 
$\6T_{(n,A)}$ into the C*-algebra of bounded $\6A_{(n,A)}$-valued 
functions on $\7R^d$.
The *-homomorphism 
\beqa
\Phi_n:\6T_{(n,A)}\rTo \9C_\8b(\7R^d,\6A_{(n,A)})
\eeqa
is given by 
\beqa
\Phi_n(f\otimes a):x \rMapsto b^{-dn^0}\sum_{x'\in b^{-n^0}\7Z^d} f(x'-x) \ 
\be_{(n,x')}(a) \ \ .
\eeqa
We introduce the C*-algebra  
\beqa
\6B_{(n,A)}&=&\8{cls}[\Phi_n(\6T_{(n,A)})]\subset \9C_\8b(\7R^d,\6A_{(n,A)})
\eeqa
The norm on $\6B_{(n,A)}$ is denoted by 
$\|\cdot\|_n$. There is a natural group homomorphism 
\beqa
\al_n\in\8{Hom}( \ \8E(d),\8{Aut}\9C_\8b(\7R^d,\6A_{(n,A)}) \ )
\eeqa  
which is given by
\beqa
(\al_{(n,g)}\1a)(y)&:=&\1a(g^{-1}y)
\eeqa
and the  *-homomorphism $\Phi_n$ is euclidean covariant
\beqa
\al_{(n,g)}\circ \Phi_n&=&\Phi_n\circ\tau_{(n,g)} \ \ .
\eeqa
In particular we obtain for $x\in b^{-n^0}\7Z^d\subset\8E(d)$:
\beqa
\al_{(n,x)}\Phi_n(f\otimes a)&:=&\Phi_n(f\otimes \beta_{(n,x)}a) \ \ .
\eeqa

\bpro\label{protra}
The translation group $\7R^d$ acts norm continuously on $\6B_{(n,A)}$
via $\al_n$.
\epro
\bpr
It is sufficient to test the continuity on the generators $\Phi_n(f\otimes a)$.
We compute 
\beqa
&&\|\Phi_n(f\otimes a)-\al_{(n,x)}\Phi_n(f\otimes a)\|_n
\vs\vs
&=&
\sup_{y\in\7R^d}\biggm\|
b^{-dn^0}\sum_{y'\in b^{-n^0}\7Z^d} 
(f(y'-y)-f(y'-y-x))\be_{(n,y')}a\biggm\|_{\6A_{(n,A)}}
\vs\vs
&\leq & b^{-dn^0}\sum_{y'\in b^{-n^0}\7Z^d} 
\sup_{y\in\7R^d}|f(y'-y)-f(y'-y-x)| \|a\|_{\6A_{(n,A)}}
\eeqa
and since $f\in S(\7R^d)$ we conclude 
\beqa
\lim_{x\to 0}\sup_{y\in\7R^d}|f(y'-y)-f(y'-y-x)| &=& 0
\eeqa
and therefore 
\beqa
\lim_{x\to 0}\|\Phi_n(f\otimes a)-\al_{(n,x)}\Phi_n(f\otimes a)\|_n &=&0
\eeqa
which proves the proposition.
\epr

Instead of the C*-inductive limit $\6A_{(\iota,A)}$, we consider 
another C*-algebra in order to build continuum limits. 
We define $\6B_{(\iota,A)}$ to be the C*-subalgebra
in $\9C_\8a(\7Z^2,\6B_A)$ which is generated by elements 
of the form 
\beqa
\Phi_{(\iota,n)}(f\otimes a)&:=&
\1p[k\mapsto \Phi_{n+k}(f\otimes\iota_{(n+k,n)}(a))] \ \ .
\eeqa
with  $a\in\6A_{(n,A)}$ and $n\in\7Z^2$. The notion 
of local algebras $\6B_{(\iota,A)}(\9U)$ is obvious. We obtain a 
euclidean net of C*-algebras $(\ul{\6B}_{(\iota,A)},\al)$ 
where the net is given by  
\beqa
\ul{\6B}_{(\iota,A)}:\9U\rMapsto \6B_{(\iota,A)}(\9U)
\eeqa
and the euclidean group acts on $\6B_{(\iota,A)}$ as follows:
\beqa
\al_g\1p[n\mapsto \1a_n]&:=&\1p[n\mapsto \al_{(n,g)}\1a_n] \ \ .
\eeqa
As a consequence of Proposition \ref{protra} we get:
\bcor
The pair $(\ul{\6B}_{(\iota,A)},\al)$, where  
$\ul{\6B}_{(\iota,A)}$ is the net 
\beqa
\ul{\6B}_{(\iota,A)}:\9U\rMapsto \6B_{(\iota,A)}(\9U)
\eeqa
is a euclidean net of C*-algebras and the translation group acts 
norm continuously on $\6B_{(\iota,A)}$.
\ecor

\subsection{On the regularity condition for continuum limits}
\label{su11}
We denote by $\hat\6S_{(\iota,A)}$ the set of states 
on $\6B_{(\iota,A)}$ such that the triple 
$(\6B_{(\iota,A)},\al,\om)$ is a weak euclidean statistical mechanics, 
i.e. it fulfills the axioms 
{\it WE1} to {\it WE3}, given in the introduction.

\bthe\label{contlim2}
There is a canonical convex-linear map 
\beqa
\1F:\6S[\9C_\8a(\7Z^2,\7C)]\otimes\Gam(\7Z^2,\6S_A)
\rTo \hat\6S_{(\iota,A)} \ \ .
\eeqa
\ethe
\bpr
For a state $\xi\in \6S[\9C_\8a(\7Z^2,\7C)]$ and a section 
$\eta\in\Gam(\7Z^2,\6S_A)$ we define $\1F[\xi\otimes\eta]$ 
by 
\beqa
\<\1F[\xi\otimes\eta],\1p[n\mapsto \1a_n]\>
&:=&
\<\xi,\1p[ \ n\mapsto \ \<\eta_n,\1a_n(0)\> \ ]\> \ \ .
\eeqa
where $\1a_n$ is contained in $\9C_\8b(\7R^d,\6A_{(n,A)})$.
It is obvious that $\1F$ is convex linear. 
In order to prove the translation invariance, 
we consider the correlation function
\beqa
\< \ \1F[\xi\otimes\eta] \ , 
\ \prod_{j=1}^k \Phi_{(\iota,n_j)}(f_j\otimes a_j) \ \>
\ \ \  = \ \ \ 
\bla\xi,\1p\biggl[ \ n \mapsto \ b^{-dn^0}
\sum_{x_1\cdots x_k\in b^{-n^0}\7Z^d} \ 
\vs\vs
\times \ \ f_1(x_1)\cdots f_k(x_k)
\ \<\eta_n , \beta_{(n,x_1)}\iota_{(n,n_1)}(a_1)
\cdots\beta_{(n,x_k)}\iota_{(n,n_k)}(a_k)\>\biggr] \bra
\eeqa
Since each $\eta_n$ is $b^{-n^0}\7Z^d$-invariant, we conclude
for $x\in\7Q_b^d$:
\beqa
&&\bla \ \1F[\xi\otimes\eta] \ , \al_x\biggl[ 
\ \prod_{j=1}^k \Phi_{(\iota,n_j)}(f_j\otimes a_j)\biggr] \ \bra
\vs\vs
&=&
\< \ \1F[\xi\otimes\eta] \ ,  
\ \prod_{j=1}^k \Phi_{(\iota,n_j)}(\tau_xf_j\otimes a_j) \ \>
\vs\vs
&=&
\bla\xi,\1p\biggl[ \ n \mapsto \ b^{-dn^0}
\sum_{x_1\cdots x_k\in b^{-n^0}\7Z^d} \ \ f_1(x_1-x)\cdots f_k(x_k-x)
\vs\vs
&\times&
\<\eta_n , \beta_{(n,x_1)}\iota_{(n,n_1)}(a_1)
\cdots\beta_{(n,x_k)}\iota_{(n,n_k)}(a_k)\>\biggr] \bra
\vs\vs
&=&
\bla\xi,\1p\biggl[ \ n \mapsto \ b^{-dn^0}
\sum_{x_1\cdots x_k\in b^{-n^0}\7Z^d} \ \ f_1(x_1)\cdots f_k(x_k)
\vs\vs
&\times&
\<\eta_n , \beta_{(n,x)}[\beta_{(n,x_1)}\iota_{(n,n_1)}(a_1)
\cdots\beta_{(n,x_k)}\iota_{(n,n_k)}(a_k)]\>\biggr] \bra
\vs\vs
&=&
\bla\xi,\1p\biggl[ \ n \mapsto \ b^{-dn^0}
\sum_{x_1\cdots x_k\in b^{-n^0}\7Z^d} \ \ f_1(x_1)\cdots f_k(x_k)
\vs\vs
&\times&
\<\eta_n ,\beta_{(n,x_1)}\iota_{(n,n_1)}(a_1)
\cdots\beta_{(n,x_k)}\iota_{(n,n_k)}(a_k) \>\biggr] \bra
\vs\vs
&=&
\< \ \1F[\xi\otimes\eta] \ ,  
\ \prod_{j=1}^k \Phi_{(\iota,n_j)}(f_j\otimes a_j) \ \>
\eeqa
which implies that $\1F[\xi\otimes\eta]$ is invariant under the 
dense subgroup $\7Q^d_b$. Since the translation group acts 
norm continuously on $\6B_{(\iota,A)}$ the states 
$\1F[\xi\otimes\eta]$ are invariant under the full translation 
group $\7R^d$. In particular, the map
\beqa
x\rMapsto \< \1F[\xi\otimes\eta] , \1a \ \al_x(\1b)\ \1c \>
\eeqa
is continuous for every $\1a,\1b,\1c\in\6B_{(\iota,A)}$.
Hence we have proven {\em WE1} and {\em WE3}. 
Let $n\mapsto \1a_n\in\6B_{(n,A)}$ be a representative of 
$\1a=\1p[n\mapsto \1a_n]$. If $\1a$ is localized in 
$\7R_+e_k+\Sgm_{e_k}$, then $\1a_n(0)$ is contained in 
$\6A_{(n,A)}(\mu,0)\otimes\6A_{(n,A)}(\mu,+)$, for $n$ large 
enough. This implies
\beqa
\< \1F[\xi\otimes\eta] , j_k(\1a) \ \1a \>
&=&
\< \xi, \1p[n\mapsto \<\eta_n,j_{(\mu,n)}(\1a_n(0))\1a_n(0) \>]\>
\vs\vs
&\geq& 0
\eeqa
according to the reflexion positivity of the $\eta_n$s. 
Thus {\em WE2} follows and the triple 
$(\6B_{(\iota,A)},\al,\1F[\xi\otimes\eta])$ is a weak 
euclidean field. 
\epr

\paragraph{Remark:}
For each section $\eta$ we introduce the set of continuum limits
\beqa
\hat\6S_{(\iota,A)}[\eta]&:=&\{\1F[\xi\otimes\eta] \ | \ 
\xi\in \6S[\9C_\8a(\7Z^2,\7C)]\}
\subset\hat\6S_{(\iota,A)} \ \ .
\eeqa
The best situation is present if 
the block spin  transformations $\iota$ are  
arranged in such a way that 
the group $\7Q_b^d$ acts norm continuously on $\6A_{(\iota,A)}$.
In this case the investigation of the set of continuum limits
$\hat\6S_{(\iota,A)}[\eta]$ on $\6B_{(\iota,A)}$ is equivalent to 
the investigation of the set of continuum limits $\6S_{(\iota,A)}[\eta]$ 
on the C*-inductive limit algebra $\6A_{(\iota,A)}$. 
Since then we conclude for the correlation function
\beqa
&&\< \ \1F[\xi\otimes\eta] \ ,  
\ \prod_{j=1}^k \Phi_{(\iota,n_j)}(f_j\otimes a_j) \ \>
\vs\vs
&=&
\int \8dx_1\cdots\8dx_k \ \prod_{j=1}^k f_j(x_j) \ 
\bla \1E[\xi\otimes\eta] \ ,  
\ \prod_{j=1}^k \be_{(\iota,x_j)}\iota_{n_j}(a_j) \ \bra
\eeqa
and 
in particular we obtain for a consistent section $\eta\in\6S_{(\iota,A)}$:
\beqa
&&\< \ \1F[\xi\otimes\eta] \ ,  
\ \prod_{j=1}^k \Phi_{(\iota,n_j)}(f_j\otimes a_j) \ \>
\vs\vs
&=&
\int \8dx_1\cdots\8dx_k \ \prod_{j=1}^k f_j(x_j) \ 
\bla \eta ,  
\ \prod_{j=1}^k \be_{(\iota,x_j)}\iota_{n_j}(a_j) \ \bra
\eeqa
which is independent of $\xi$. 

\section{Conclusion and outlook}
\label{s9}
\paragraph{\it Concluding remarks:}
Some of the basic ideas and concepts which are used in order 
to construct euclidean field theory models are generalized 
by using the setup of algebraic euclidean field theory. 
We have introduced the notions block spin transformations, action, and 
effective action within a general model independent framework.

As described in Section \ref{s2} and Section \ref{s3}, 
in the C*-algebraic approach to euclidean field theory 
the concept of continuum limits for lattice field theories 
arises in a very natural manner. To each section 
$\eta\in\Gam(\7Z^2,\6S_A)$, which is a family of 
lattice field theory models (these models can be chosen 
on each lattice $\Sgm_d(n)$ independently from each other), 
there always exists the corresponding set 
$\6S_{(\iota,A)}[\eta]$ of continuum limits. 

Therefore, our point of view leads to a well posed problem. 
In order to prove the existence of non-trivial (weak) 
euclidean field theory 
models, one has to study the properties 
of the set of continuum limits with respect to the 
properties of the corresponding section $\eta$.

\paragraph{\it Outlook:}
It would be desirable to study the continuum limits, which arise from 
lattice models with an action (see Section \ref{s3}) of the form   
\beqa
\1v[h]_n&:=&\prod_{\Gam\in\Sgm_{d-1}(n)} \Phi_n(\Gam,w)
\ \ ,
\eeqa 
in more detail. As already mentioned in the introduction, one of the   
questions, which we want to investigate, 
is the following:

\paragraph{\it Question:} Which are sufficient conditions for 
the family of operators $w=(w_n)_{n\in\7Z^2}\subset M'\olt M'$ such that 
the set of continuum limits $\6S_{(\iota,A)}[\eta]$

\bdes
\itno 1 contains only characters (in case of abelian C*-algebras)?

\itno 2 contains only ultra local states?

\itno 3 contains at least one state which is not ultra local? 
\edes

The states $\varphi \in\6S_{(\iota,A)}[\eta] $ are weak 
limit points and labeled by states $\xi$ on the 
corona algebra $\9C_\8a(\7Z^2,\7C)$. The states 
$\xi$ are not explicitly given, namely its existence is related 
to the Hahn-Banach extension theorem and therefore it relies on 
Zorn's lemma, however. In order to conclude properties for 
the continuum limits one has to think about which type 
of statements one can prove. For instance,
one can use operators in $\6A_{(n,M)}$ to test properties of the 
states $\eta_n$ like bounds of correlation functions.

In order to decide whether case {\it (3)} is valid, we propose to 
compute correlations 
\beqa
\<\1c_{[\eta_{n+k}\circ\iota_{(n+k,n)}]},
\Phi_n(\Delta_1,a)\otimes\Phi_n(\Delta_2,a)\>
\eeqa
for an appropriate choice of the operator $a>0$.
Then one has to arrange each operator $w_n$ in such a way that the 
bound
\beqa
|\<\1c_{[\eta_{n+k}\circ\iota_{(n+k,n)}]},
\Phi_n(\Delta_1,a)\otimes\Phi_n(\Delta_2,a)\>|
&>&c_{(n,\Delta_1,\Delta_2,a)}
\eeqa
is fulfilled with a positive constant 
$c_{(n,\Delta_1,\Delta_2,a)}$ which only depends on $n$ 
the cubes $(\Delta_1,\Delta_2)$ and the operator $a$. 
Within Appendix \ref{app1}, we discuss a strategy how to 
deal with this problem.  

In Section \ref{s3} the notion of effective action for continuum limits is 
discussed. Let $(X,\9P,\om_o)$ be a measure space with $\sgm$-algebra 
$\9P$ and we consider the von Neumann algebra $M=\9L^\infty(X,\9P,\om_o)$
and the states $\om_n:=\otimes_{\Delta\in\Sgm_d(n)}\{\Delta,\om_o\}$.
Let $\varphi\in\6S_{(\iota,M)}[\eta]$ be a continuum limit for which 
the effective action $\1v$ exists, i.e.
\beqa
\<\varphi,\iota_na\>&=&\int \8d\om_n \ \1v_n \ a \ \ .
\eeqa 
Then $\1v_n$ is a $\otimes_{\Delta\in\Sgm_d(n)}\9P$-measurable
function.   

Let $X$ be a smooth orientable manifold, let $\9P$ be the 
$\sgm$-Borel algebra and let $\om_o$ be a volume form on $X$, then one can 
ask for a criterion for the section $\eta$ such that 
the effective action $\1v$ is a section of smooth functions $\1v_n$ on 
$X^{\Sgm_d(n)}$. Within a coordinate chart $(\phi^\sgm)_{\sgm=0,\cdots,p}$, 
at $u_o\in X$ one can perform a 
Taylor expansion of the effective action functional 
$\1s_n=-\ln\1v_n$ at $u_{(o,n)}:\Delta\mapsto u_o$ 
\beqa
\1s_n&=& 
\sum_{k=0}^K {1\over k!} \sum_{(\Delta_j,\sgm_j)} 
\phi^{(\Delta_1,\sgm_1)}\cdots\phi^{(\Delta_k,\sgm_k)}
\pa_{(\Delta_1,\sgm_1)}\cdots\pa_{(\Delta_k,\sgm_k)}\1s_n(u_{(o,n)})
\vs\vs
&+&\8{reminder}
\eeqa
where $(\phi^{(\Delta,\sgm)})_{\Delta\in\Sgm_d(n),\sgm=0,\cdots,p}$ 
is the coordinate chart of $X^{\Sgm_d(n)}$
induced by $(\phi^\sgm)_{\sgm=0,\cdots,p}$. The {\em free part} $\1v^{(0)}$
of $\1v$ can be defined by 
\beqa
\1v^{(0)}_n &=&  \exp[-\<\phi,\1A_n\phi\>]
\eeqa
where the quadratic form $\1A_n$ is given by 
\beqa
\<\phi,\1A_n\phi\>&=& {1\over 2}
\sum_{(\Delta_1,\sgm_1),(\Delta_2,\sgm_2)} 
\phi^{(\Delta_1,\sgm_1)}\phi^{(\Delta_2,\sgm_2)}
\pa_{(\Delta_1,\sgm_1)}\pa_{(\Delta_2,\sgm_2)}\1s_n(u_0) \ \ .
\eeqa 
Since the sum over the 
pairs $(\Delta_1,\sgm_1),(\Delta_2,\sgm_2)$ may also contain 
cubes $(\Delta_1,\Delta_2)$ which are not next neighbors, 
we expect that in general $\1v^{(0)}$ is not an action. 

Nevertheless, it makes sense to study the section of 
gaussian states $\eta^{(0)}$, where $\eta^{(0)}$ is a state 
on $\6A_{(n,T_{u_o}M)}$ and $T_{u_o}M$ is the von Neumann
algebra $\9L^\infty(T_{u_o}X)$ of Lebesgue measurable functions
on the tangent space $T_{u_o}X$ at $u_o$.
If we assume that $\1A_n$ is a positive quadratic form, then 
we obtain for the characteristic functional 
\beqa
\<\eta^{(0)}_n,\exp(\phi(f))\>&=&\exp( -\<f,\1A_n^{-1}f\>) \ \ .
\eeqa
This implies that the continuum limits in 
$\6S_{(\iota,T_{u_o}M)}[\eta^{(0)}]$ 
are (mixtures of) gaussian states. We propose to  compare the set of 
continuum limits \newline
$\6S_{(\iota,T_{u_o}M)}[\eta^{(0)}]$ with the set of  
continuum limits $\6S_{(\iota,M)}[\eta]$ of the underlying section $\eta$ 
in order to decide whether there are states $\varphi \in 
\6S_{(\iota,M)}[\eta]$ which describe a physical system 
with interaction phenomena.

\subsubsection*{{\it Acknowledgment:}}
I am grateful to Prof. Jakob Yngvason for 
supporting this investigation with hints and many ideas.
This investigation is financially supported by the 
Deutsche Forschungsgemeinschaft (DFG) which is also gratefully acknowledged.
Finally I would like to thank the 
Erwin Schr\"odinger International Institute for Mathematical Physics, 
Vienna (ESI) for its hospitality.
\newpage
\begin{appendix}
\section{Criterion for the existence of non ultra local continuum limits}
\label{app1}
We use the following notation: We choose a von Neumann algebra 
$M$ acting on $K$  and a cyclic and separating vector $\Om\in K$  
and $\om $ is the state $\om:=\<\Om,(\cdot)\Om\>$.
For a positive operator $w\in M'\olt M'$
$\Delta_1,\Delta_2\in\Sgm_d(n)$ we introduce a 
correlation functional on $M\olt M$ by 
\beqa
\<\1c_{(\om,w,n)}^{(\Delta_1,\Delta_2)},a_1\otimes a_2\>
&:=&
\<\eta_{(\om,w,n)},\Phi_n(\Delta_1,a_1)\Phi_n(\Delta_2,a_2)\>
\vs\vs
&-&
\<\eta_{(\om,w,n)},\Phi_n(\Delta_1,a_1)\>
\<\eta_{(\om,w,n)},\Phi_n(\Delta_2,a_2)\>
\eeqa
where $\eta_{(\om,w,n)}$ is the state which is given by 
\beqa
\<\eta_{(\om,w,n)},a\>&:=&\1z_{(\om,w,n)}^{-1}
\bla \eta_{(\om,n)}, 
\prod_{\Gam\in\Sgm_{d-1}(n)}\Phi_n(\Gam,w) \ a \bra \ \ .
\eeqa

We now introduce particular classes of positive 
operators in $M'\olt M'$.

\bdef
For a constant $2 > c > 0$ and cubes $\Delta_1,\Delta_2\in\Sgm_d(n)$
and projections $P_1,P_2\in\8{Proj}(M)$ we define the set 
\beqa
\9P_{(c,n)}^{[\Delta_1,\Delta_2;P_1,P_2]}&:=&\biggl\{w\in M'\olt M'\biggm| 
\vs\vs
&& w>0 \ \mbox{ and } \  
|\<\1c_{(\om,w,n)}^{[\Delta_1,\Delta_2]},P_1\otimes P_2\>| > c \biggr\} \ \ .
\eeqa
\eef

\paragraph{Remark:}
For each translation $g\in b^{-n_0}\7Z^d$ we obtain the identity
\beqa
\9P_{(c,n)}^{[g\Delta_1,g\Delta_2;P_1,P_2]}&=&
\9P_{(c,n)}^{[\Delta_1,\Delta_2;P_1,P_2]} \ \ .
\eeqa
Furthermore, we have $\9P_{(c,n)}^{[\Delta_1,\Delta_2;P,\11]}=\emptyset$
for each projection $P\in\8{Proj}(M)$.
\abs
 
One easily computes the relation
\beqa
\<\1c_{(\om,w,n)}^{(\Delta_1,\Delta_2)},(\11-P_1)\otimes P_2\>
&=&
-\<\1c_{(\om,w,n)}^{(\Delta_1,\Delta_2)},P_1\otimes P_2\>
\eeqa
and therefore 
\beqa
\<\1c_{(\om,w,n)}^{(\Delta_1,\Delta_2)},P_1\otimes P_2\>
&=&
\<\1c_{(\om,w,n)}^{(\Delta_1,\Delta_2)},(\11-P_1)\otimes (\11-P_2)\>
\eeqa
and we obtain the identity
\beqa
\9P_{(c,n)}^{[\Delta_1,\Delta_2;P_1,P_2]} &=&  
\9P_{(c,n)}^{[\Delta_1,\Delta_2;\11-P_1,P_2]}
\vs\vs
&=&
\9P_{(c,n)}^{[\Delta_1,\Delta_2;\11-P_1,P_2]}
\vs\vs
&=&
\9P_{(c,n)}^{[\Delta_1,\Delta_2;\11-P_1,\11-P_2]}
\eeqa

By using the block spin transformation introduced in Section \ref{su5}
we obtain  
\beqa
\iota_{(n+k,n)}\Phi_n(\Delta,P)&=&\Phi_{n+k}(\Delta_{(n+k,n|\Delta)},P)
\eeqa
and we may define for each $k\in\7N^2$ the sets 
\beqa
\9P_{(c,n+k,n)}^{[\Delta_1,\Delta_2;P_1,P_2]}&:=&
\9P_{(c,n)}^{[\Delta_{(n+k,n|\Delta_1)},\Delta_{(n+k,n|\Delta_2)};P_1,P_2]} 
\ \ .
\eeqa

According to Proposition \ref{proact}, a section 
\beq\label{eqa1}
w:k\rMapsto w_k\in\9P_{(c,n+k,n)}^{[\Delta_1,\Delta_2;P_1,P_2]}
\eeq
yields an action $\1v[h]$ by 
\beqa
\1v[w]_{n+k}&=&
\prod_{\Gam\in\Sgm_{d-1}(n+k)}\Phi_{n+k}(\Gam,w_k) 
\eeqa
and therefore a section of reflexion positive invariant states
$\eta$. 

\bpro
Let $w$ be a section, given by Equation (\ref{eqa1}) and let 
$\eta$ be the corresponding section of reflexion positive invariant states.
Then the set of continuum limits $\6S_{(\iota,A)}[\eta]$
contains at least one state which is not ultra local, i.e.
case {\it (3)} is valid (see Introduction).
\epro
\bpr
According to the definition of $\9P_{(c,n+k,n)}^{[\Delta_1,\Delta_2;P_1,P_2]}$
we obtain the bound 
\beqa
c &<&
|\<\1c_{[\eta_{n+k}\circ\iota_{(n+k,n)}]}, \Phi_n(\Delta_1,P_1)\otimes
\Phi_n(\Delta_2,P_2)\> | \ \ .
\eeqa
We define the subset $\7X_+\subset\7Z^2$ to consist of all 
$n_1\in\7Z^2$ such that $n\prec n_1$ and 
\beqa
c &<&
\<\1c_{[\eta_{n_1}\circ\iota_{(n_1,n)}]}, \Phi_n(\Delta_1,P_1)\otimes
\Phi_n(\Delta_2,P_2)\>
\eeqa
and write $\7X_-:=\7Z^2\bs\7X_+$. Then there exists 
a character 
\beqa
\xi\in\6S[\9C_\8a(\7X_+,\7C)]\cup\6S[\9C_\8a(\7X_-,\7C)]
\subset\6S[\9C_\8a(\7Z^2,\7C)]
\eeqa
which implies 
\beqa
c &<&|\<\1c_{[\1E[\xi\otimes\eta]_n]}, \Phi_n(\Delta_1,P_1)\otimes
\Phi_n(\Delta_2,P_2)\> |
\eeqa
and the state $\1E[\xi\otimes\eta]$ is not ultra local.
\epr

In order to prove the existence of non-ultra local continuum limits 
one has to check the assumption of the following corollary:
\bcor
If for a projection $P\in\8{Proj}(M)\bs\{\11\}$ and 
for each pair of cubes $\Delta_1,\Delta_2\in\Sgm_d(n)$
there exists a constant $c[\Delta_1,\Delta_2,P]\in [2,0)$ such that 
\beqa
\9P_{(c[\Delta_1,\Delta_2,P],n)}^{[\Delta_1,\Delta_2;P,P]}
&\not=&\emptyset \ \ ,
\eeqa
then there exists 
a non ultra local state in $\6S_{(\iota,M)}$.
\ecor

\section{Multiplicatively renormalizable actions: An example}
\label{app2}
We consider the von Neumann algebra $M=\9L^\infty([0,1])$ and  
a family of positive functions $h\in\9C^\infty([0,1]^2)^{\7Z^2}$.
The action of the model under consideration is 
given by 
\beqa
\1v[h]_n(u)&:=&\int \prod_{\Gam\in\Sgm_{d-1}(n)} \8ds(\Gam) \ 
\prod_{\Delta\in\Sgm_d(n)} \ \prod_{\Gam\in\pa\Delta}h_n(u(\Delta),s(\Gam))  
\eeqa
and we introduce the function
\beqa
\1H^{(\Delta,I)}_{(h,n)}(s)&:=&\int_{I}
\8du \ \prod_{\Gam\in\pa\Delta}h_n(u,s(\Gam)) 
\eeqa
Let $y_n\in\9C^\infty(\7R)$ be a smooth positive function 
with $y_n(s)\geq 1$ for each $s$. 
We choose $h_n(u,s):=\exp(uy_n(s))$ and by setting 
\beqa
\1y_{(n,\Delta)}(s)&:=&\sum_{\Gam\in\pa\Delta}y(s(\Gam))
\eeqa
we obtain:
\beqa
\1H^{(\Delta,[u_0,u_1])}_{(h,n)}(s)&=&
\1y_{(n,\Delta)}(s)^{-1}
[\exp(u_1\1y_{(n,\Delta)}(s))-\exp(u_0\1y_{(n,\Delta)}(s))] \ \ .
\eeqa
By introducing $\1q_n:=\sup_s\1y_{(n,\Delta)}(s)$ and 
$\1r_n:=\inf_s\1y_{(n,\Delta)}(s)$ we conclude:
\beqa
S_{(\om,n)}(h)&=&\1q_n^{-1}(\exp(\1q_n)-1)
\vs\vs
I_{(\om,n)}(h)&=&\1r_n^{-1}(\exp(\1r_n)-1) \ \ .
\eeqa
and for each $k\in\7N^2$ we get 
\beqa
\biggl[{S_{(\om,n+k)}(h)\over I_{(\om,n+k)}(h)}\biggr]^{\tau(n+k)}
S_{(\om,n+k)}(h)^{-\tau(n)}
&=&
\biggl[{\1r_{n+k}\over\1q_{n+k}}\biggr]^{\tau(n+k)}
\vs\vs
&\times&
\biggl[{\exp(\1q_{n+k})-1\over\exp(\1r_{n+k})-1}\biggr]^{\tau(n+k)}
\vs\vs 
&\times&
\1q_{n+k}^{\tau(n)}(\exp(\1q_{n+k})-1)^{-\tau(n)} \ \ .
\eeqa

The action $\1v[h]$ is multiplicatively renornmalizable if  
the values of $\1q_n$ and $\1r_n$ can be arranged in such a 
way that the following holds true: 
\bdes
\itno 1
There exists a constant $\1c>1$ such that 
\beqa
\1c:=\lim_{n\in\7Z^2}\1q_n=\lim_{n\in\7Z^2}\1r_n \ \ .
\eeqa

\itno 2
The supreme 
\beqa
\1S_n&:=&\sup_{k\in\7N^2}
\biggl[{\exp(\1q_{n+k})-1\over\exp(\1r_{n+k})-1}\biggr]^{\tau(n+k)}
\eeqa
is finite for each $n\in\7Z^2$.
\edes
Then one easily computes 
\beqa
1 \  \ \leq \ \ [[ \ \1r_\om\1v[h] \ ]]_n&\leq& \8{const.} \ 
\1S_n\1c^{\tau(n)}(\exp(\1c)-1)^{-\tau(n)}  \ \ .
\eeqa
\end{appendix}
\newpage



\begin{thebibliography}{References}

\bibitem{AshLew95}
Ashtekar, A. and Lewandowski, J.:\\
{\it Differential geometry on the space of connections via graphs 
and projective limits.}\\
J. Geom. Phys. {\bf 17}, 191-230, (1995)

\bibitem{Bal84a}
Balaban, T.:\\
{\it Propagators and renormalization group transforms 
for triangulation gauge theories I.}\\
Commun. Math. Phys. {\bf 95}, 17-40, (1984) 

{\it Propagators and renormalization group transforms 
for lattice gauge theories II.}\\
Commun. Math. Phys. {\bf 96}, 223-250, (1984)

\bibitem{BrDiHu97}
Brydges, D., Dimock, J. and Hurd, T. R.:\\
{\it Estimates on renormalization group transformations.}\\
(1997)

\bibitem{BuVer97}
Buchholz, D. and Verch, R.:\\
{\it Scaling algebras and renormalization group in algebraic 
quantum field theory. II. Instructive examples}\\ 
hep-th/9708095

\bibitem{BuVer95}
Buchholz, D. and Verch, R.:\\
{\it Scaling algebras and renormalization group in algebraic 
quantum field theory}\\ 
Rev. Math. Phys. {\bf 7}, 1195-1240, (1995) 

\bibitem{DrieFroh77}
Driessler,W. and Fr\"ohlich, J.:\\
{\it The reconstruction of local algebras from the 
Euclidean Green's functions of relativistic quantum field theory.}
Ann. Inst. Henri Poincar\'e {\bf 27}, 221-236, (1977)

\bibitem{FeldOst}
Feldman, J. and Osterwalder, K.:\\
{\it The Wightman axioms and the mass gap for weakly coupled 
$\phi^4_3$ quantum field theories.}\\
In: Mathematical Problems in Theoretical Physics, H. Araki, ed.,
Berlin, Heidelberg, new York: Springer-Verlag.

\bibitem{Froh79}
Fr\"ohlich, J.:\\
{\it Some results and comments on quantized gauge fields.}\\ 
Cargese, Proceedings, Recent Developments In Gauge Theories, 53-82, 
(1979)

\bibitem{FernFrohSok92}
Fern\'andez, R.,Fr\"ohlich, J. and Sokal, A. D.:\\
{\it Random walks, critical phenomena, and trivialty in 
quantum field theory.}\\
Berlin, Heidelberg, New York: Springer-Verlag. (1992) 

\bibitem{Galnico85}
Gallavotti, G. and nicol\`o, F.:\\
{\it Renormalization theory in four-dimensional 
scalar fields I.}\\
Commun. Math. Phys. {\bf 100}, 545-590, (1985)
\\
{\it Renormalization theory in four-dimensional 
scalar fields II.}\\
Commun. Math. Phys. {\bf 101}, 247-282, (1985)

\bibitem{GalRiv84}
Gallavotti, G. and Rivasseau, V.:\\
{\it $\Phi^4$ field theory in dimension 4: a modern introduction to its 
unsolved problems.}\\
Ann. Inst. Henri Poincar\'e {\bf 40}, 185-220, (1984)

\bibitem{GawKup}
Gawedzki, G. and Kupiainen, A.:\\
{\it Asymptotic freedom beyond perturbation theory.}\\
Les Houches lectures (1984)

\bibitem{GlJa1}
Glimm, J. and Jaffe, A.:\\ 
{\it Collected Papers. Vol. 1 and Vol.2: 
Quantum field therory and statistical mechanics.}\\
Expositions.
Boston, USA: Birkh\"auser (1985) 

\bibitem{GlJa3}
Glimm, J. and Jaffe, A.:\\ 
{\it The Yukawa-2 quantum field theory witohut cutoffs.}\\
J. Funct. Anal. {\bf 7}, 323-357, (1971)

\bibitem{GlJa5}
Glimm, J. and Jaffe, A.: \\
{\it Positivity of the $\phi^4_3$ hamiltonian.}\\
Fortschritte der Physik {\bf 21}, 327-376, (1973)

\bibitem{GlJa4}
Glimm, J. and Jaffe, A.: \\
{\it Quantum physics, a functional integral point of view.}\\
Springer, new York, Berlin, Heidelberg (1987)

\bibitem{HK} 
Haag, R. and Kastler, D.:\\ 
{\it An algebraic approach to quantum field theory.}\\
J. Math. Phys. {\bf 5}, 848-861, (1964)

\bibitem{King86}
King, C.:\\
{\it The U(1) higgs model II. The infinite volume limit.}\\
Commun. Math. Phys. {\bf 103}, 323-349, (1986)

\bibitem{Klau70}
Klauder, J. R.: \\
{\it Ultralocal scalar field models.}\\
Commun. Math. Phys. {\bf 18}, 307-318, (1970) 

\bibitem{MagRivSen93}
Magnen, J., Rivasseau, V. and S\'en\'eor, R.:\\
{\it Construction of YM-4 with an infrared cutoff.}\\
Commun. Math. Phys. {\bf 155}, 325-384, (1993) 

\bibitem{MagSen76}
Magnen, J. and S\'en\'eor, R.:\\
{\it The infinite volume limit of the $\phi^4_3$ model.}\\
Inst. H. Poincar\'e {\bf 24}, 95-159, (1976)

\bibitem{OstSchra1}
Osterwalder, K. and Schrader, R.:\\ 
{\it Axioms for Euclidean Green's functions I.}\\
Commun. Math. Phys. {\bf 31}, 83-112, (1973)

Osterwalder, K. and Schrader, R.:\\ 
{\it Axioms for Euclidean Green's functions II.}\\
Commun. Math. Phys. {\bf 42}, 281-305, (1975)


\bibitem{Roo}
Roos, H.:\\ 
{\it Independence of local algebras in quantum field theory}\\
Commun. Math. Phys. {\bf 16}, 238-246, (1970)

\bibitem{Schl97}
Schlingemann, D.:\\
{\it From euclidean field theory to quantum field theory.}\\
To appear in Rev. Math. Phys. (1998)

\bibitem{Schra0}
Schrader, R.:\\ 
{\it A remark on Yukawa plus boson self-interaction in two space-time
dimensions.}\\ 
Commun. Math. Phys. {\bf 21}, 164-170, (1971) 

\bibitem{Schra1}
Schrader, R.:\\ 
{\it A Yukawa quantum  field theory in two space-time dimensions 
without cutoffs.}\\ 
Ann. Phys.  {\bf 70}, 412-457, (1972)

\bibitem{Seil82}
Seiler, E.:\\
{\it Gauge theories as a problem of constructive quantum field theory
and statistical mechanics.}\\
Berlin, Germany: Springer (1982) 192 P. ( Lecture notes In Physics, 159). 

\bibitem{SeilSim76}
Seiler, E. and  Simon, B.:\\
{\it Nelson's symmetry and all that in the Yukawa$_2$ and 
$\phi^4_3$ field theories.}\\
Ann. Phys. {\bf 97}, 470-518, (1976)


\end{thebibliography}
\end{document}